\documentclass[citeautoscript,floatfix,aps,prx,twocolumn,superscriptaddress,longbibliography]{revtex4-1}  
\usepackage{graphicx}  
\usepackage{dcolumn}   
\usepackage{bm}        
\usepackage{amssymb}   
\usepackage{amsmath}
\usepackage{amsthm}
\newtheorem{theorem}{Theorem}
\usepackage{multirow}
\usepackage{natbib}

\DeclareMathOperator{\arcsinh}{arcsinh}

\usepackage{braket}
\usepackage{amsfonts}
\usepackage{dsfont}
\usepackage{hyperref}
\usepackage{xcolor}
\newcommand{\ts}{\textsuperscript}
\usepackage{mathtools}

\begin{document}

\title{Tightening the Lieb-Robinson Bound in Locally-Interacting Systems}
\author{Zhiyuan Wang}
\affiliation{Department of Physics and Astronomy, Rice University, Houston, Texas 77005,
  USA}
\affiliation{Rice Center for Quantum Materials, Rice University, Houston, Texas 77005, USA}
\author{Kaden R.~A. Hazzard}
\affiliation{Department of Physics and Astronomy, Rice University, Houston, Texas 77005,
USA}
\affiliation{Rice Center for Quantum Materials, Rice University, Houston, Texas 77005, USA}
\date{\today}

\begin{abstract}
  The Lieb-Robinson~(LR) bound rigorously shows that in quantum systems with short-range interactions, the maximum amount of information that travels beyond an effective ``light cone'' decays exponentially with distance from the light-cone front, which expands at finite velocity. Despite being a fundamental result, existing bounds are often extremely loose, limiting their applications. We introduce a method that dramatically and qualitatively improves LR bounds in models with finite-range interactions. Most prominently, in systems with a large local Hilbert space dimension $D$, our method gives an LR velocity that grows much slower than previous bounds with $D$ as $D\to \infty$. For example,
  in the Heisenberg model with spin $S$, we find $v\leq$~const. compared to the previous $v\propto S$ which diverges at large $S$, 
  and in multiorbital Hubbard models with $N$ orbitals, we find $v\propto \sqrt{N}$ instead of previous $v\propto N$, and similarly in the $N$-state truncated Bose-Hubbard model and Wen's quantum rotor model. Our bounds also scale qualitatively better in some systems when the spatial dimension or certain model parameters become large, for example in the $d$-dimensional quantum Ising model and perturbed toric code models. Even in spin-1/2 Ising and Fermi-Hubbard models, our method improves the LR velocity by an order of magnitude with typical model parameters, and significantly improves the LR bound at large distance and early time. 
  
\end{abstract}

\maketitle

\section{Introduction}
The Lieb-Robinson~(LR) bound~\cite{Lieb1972,hastings2010locality} has revealed how locality shapes and constrains quantum matter. It says that in short-range (exponentially decaying) interacting quantum systems on a lattice, the influence of any local perturbation is restricted to an effective light cone expanding at a finite speed, apart from an exponentially decaying tail outside the light cone. It has direct implications for many body quantum dynamics, as it has been used to understand the timescales necessary to generate correlation and entanglement~\cite{bravyi2006,eisert2006arealaw,Hazzard2014quantum,gogolin2016equilibration}, to transfer information in a quantum chanel~\cite{bose2007quantum,epstein2017quantum}, and to prethermalize a system~\cite{prethermalization1,prethermalization2,prethermalization3,thermalization}. It has also found applications to equilibrium properties, for example in proving that correlations decay exponentially in a gapped ground state~(the exponential clustering theorem~\cite{hastings2004decay,hastings2006,nachtergaele2006,cramer2006correlations,hastings2006solving,haegeman2013elementary,kliesch2014locality}), that systems with half-integer spin per unit cell are necessarily gapless~(the Lieb-Schultz-Mattis-Hastings theorem~\cite{LSM,LSMhigherD,LSMhigherD2}), the area law for entanglement entropy in ground states of gapped systems~\cite{hastings2007entropy,hastings2007area,RMP2010Arealaw}, the stability of topological order~\cite{hastings2005quasiadiabatic,bravyi2010topological,TPstability,michalakis2013stability}, and the formal proof for the quantization of the Hall conductance as integer multiples of $e^2/h$~\cite{hastings2015proofIQHE}. It is also the theoretical basis for analyzing the accuracy of some numerical algorithms~\cite{Osborne2006,Osborne2007a,Osborne2007b,woods2015simulating,woods2016dynamical,tran2019locality}. It has been generalized to systems with long-range power-law decaying interaction~\cite{hastings2006,Eisert2013breakdown,Gorshkov2014,Gong2014persistence,Foss2015nearly,maghrebi2016causality,matsuta2017improving,else2020improved,guo2019signaling,tran2019locality,Lucas2019longrange,kuwahara2020strictly}, to $n$-partite connected correlation functions~\cite{tran2017npartite}, and to dynamics with Markovian dissipation~\cite{hastings2004locality,Poulin2010,sweke2019lieb} as well. Furthermore, recent developments in experimental technology of atomic, molecular and optical physics have enabled experimentalists to observe the effective light-cone and measure the LR speed~\cite{Exp1,Exp2,Gorshkov2014}.

Despite the LR bounds' fundamental consequences and efforts to tighten them~\cite{nachtergaele2006,hastings2006,Hamma2009PRL,Poulin2010,schuch2011information,Roberts2016PRL}, the bounds are often extremely loose~(except some special bounds on free particle systems~\cite{burrell2007bounds,cramer2008locality,burrell2009information,damanik2014new,gong2019lieb}), severely limiting their applicability. 
To see this, note that in locally interacting systems, previous LR bounds are typically of the form
\begin{equation}\label{eq:previousLR}
  \|[\hat{A}_X(t),\hat{B}_Y(0)]\|\leq C e^{-s_{XY}/\xi}(e^{v|t|/\xi}-1),
\end{equation}
where $\hat{A}_X,\hat{B}_Y$ are local operators supported on regions $X,Y$, respectively, $\hat{A}_X(t)=e^{i\hat{H}t}\hat{A}_Xe^{-i\hat{H}t}$, $s_{XY}$ is the minimal graph-theoretical distance between points in $X$ and $Y$, and $v$ and $\xi$ are constants that only depend on the Hamiltonian $\hat{H}$~(and the lattice structure). 
There are two qualitative ways in which these bounds are loose. 
First, and arguably most importantly, the LR speeds $v$ appearing in existing bounds  are typically much larger than the actual speed of information propagation. For example, in the one-dimensional~(1D) transverse-field
Ising model~(TFIM) at critical point, the previous best LR speed is $8eJ\approx 21.7 J$ while the actual speed determined from exact solution is $2J$. Previous bounds on $v$ can even become infinitely loose in certain limits: for example in the TFIM at large $J$, previous bounds grow $\propto J$; in the spin-$S$ Heisenberg model, previous bounds grow $\propto S$; in the classical limit of the SU$(N)$ Fermi-Hubbard~(FH) model, previous bounds grow $\propto N$, while in all these cases the true LR speed remains finite. 
Previous bounds are also qualitatively loose at short time and large distance: as $t\to 0$, perturbative arguments show that the left-hand side~(LHS) of Eq.~\eqref{eq:previousLR} grows like $O(t^{c s_{XY}})$, where $c$ is a constant, much slower than the $O(t)$ in the RHS of Eq.~\eqref{eq:previousLR}; similarly, as $s_{XY}\to \infty$, the LHS actually decays faster than exponential. A simple example is non-interacting systems, where the LHS decays at large distance like $O(e^{-s_{XY}\ln s_{XY}})$. 

In this paper, we focus on locally interacting systems~(\textit{i.e.} systems with finite-range interactions), and introduce a  method that qualitatively tightens the LR bounds in all the aforementioned aspects. 
The key new ingredient in our method is the \textit{commutativity graph}, which is a graphical tool that helps us taking advantage of the commutativity structure of the Hamiltonian terms, a feature that has been ignored by previous methods~(except in a special model in Ref.~\cite{Hamma2009PRL}).
By applying the Heisenberg equation and triangle inequality, we obtain a system of linear integral inequalities which only involve the unequal time commutator of the Hamiltonian terms. We then find a system of linear differential equations  
whose solution gives an upper bound for any solution to the integral inequalities. These differential equations can be naturally understood as the information flow equation on the commutativity graph.

If quantitative tightness is the sole criterion for applying the LR bound, one should solve these equations either analytically, or numerically if necessary, which can be done quickly and straightforwardly for systems with many thousands of sites~[since the number of variables is proportional to the size of the system, see Eq.~\eqref{eq:dfbart}]. In translationally invariant systems, the solution to the system of linear differential equations can be reduced to an integral by Fourier transformation.
For example, in the $d$-dimensional TFIM, our  LR bound for $[\hat{\sigma}^x_{\vec{r}}(t),\hat{\sigma}^z_{\vec{0}}(0)]$ is
\begin{equation}\label{eq:fourierintegralsolu_intro}
  \|[\hat{\sigma}^x_{\vec{r}}(t),\hat{\sigma}^z_{\vec{0}}(0)]\|\leq \int^\pi_{-\pi}\cosh[\omega(\vec{k})t]e^{i\vec{k}\cdot \vec{r}}\frac{d^dk}{(2\pi)^d},
\end{equation}
where $\omega(k)$ corresponds to eigenvalue of the linear operator encoding the differential equations, which can be found by diagonalizing a $(d+1)\times (d+1)$ matrix for each $k$. The bound for other operators takes a similar form. Based on the Fourier integral representation, we provide a general method to analyze the asymptotic behaviors of the solutions--which shows that the small $t$ and large $r$ behavior is qualitatively tighter than prior bounds, and are often tightest possible--and derive an explicit formula to extract the LR speed. The LR speed obtained by this method vastly improves previous ones, as summarized in Tables~\ref{tab:comp} and \ref{tab:compqual}.

As we will see, there are three broad scenarios where the LR speed appearing in our  bounds is qualitatively tighter: (1) when the number of local degrees of freedom becomes large, e.g. the large-$S$ limit in TFIM, Heisenberg XYZ models, and Wen's quantum rotor model; the Bose-Hubbard~(BH) model truncated to allow at most $N$ particles per site, and the large-$N$ limit in SU($N$) FH model, (2) when the parameters of a set of commuting operators becomes large~(e.g. large-$J$ limit in TFIM, perturbed toric code model), and (3) TFIM and Wen's quantum rotor model in large spatial dimension.
The bounds we present shed light onto these important limits which have previously resisted bounds with the qualitatively correct scaling. They also have important physical implications, for example in exactly solvable models with topological order~(e.g., toric code~\cite{kitaev2003fault} and string net models~\cite{levin2005string}), perturbed by an arbitrary bounded, local perturbation. Our results show that their LR velocity vanishes as the strength of the perturbation approaches zero, in contrast to prior bounds which retained a finite LR velocity in this limit.
Even in situations with low spatial dimension and small local degrees of freedom, e.g. at $J=h$ in 2D TFIM or $U=J$ in 1D FH, our bound represents more than a 10-fold improvement.

Although expressions like Eq.~\eqref{eq:fourierintegralsolu_intro} provide the tightest bounds in this paper, we can find simple analytic bounds while compromising the tightness only marginally. We will
show that
\begin{equation}\label{eq:factorialLR}
  \|[\hat{A}_X(t),\hat{B}_Y(0)]\|\leq C \left(\frac{u|t|}{d_{XY}}\right)^{d_{XY}}.
\end{equation}
We note that this bound applies even to systems lacking translational invariance~(under mild realistic requirements on the Hamiltonian to be introduced later).
Here $d_{XY}$ is the distance between operators $\hat{A}_X(t),\hat{B}_Y(0)$ on the commutativity graph~($d_{XY}$ grows linearly with the distance between the two operators in real space), $C$ is a constant, and $u$ is related to the LR speed given in Tables~\ref{tab:comp},~\ref{tab:compqual}~(their precise relation is given in Sec.~\ref{sec:factbound} and Sec.~\ref{sec:examples}). This bound also tightens previous ones in the two aspects mentioned above, and the asymptotic behavior at small time and large distance is often tightest possible. For example, the small-$t$ exponent is the same as the exact one obtained by perturbative arguments and the large-$x$ exponent is saturated by some free particle systems.

 \begin{table}[ht]
  \centering
  {\renewcommand{\arraystretch}{1.5}
  \begin{tabular}{|c|l|l|l|l|}
    \hline
model   & \multicolumn{2}{c|}{$v_{\text{LR}}$~(this paper)}  & \multicolumn{2}{c|}{$v_{\text{LR}}$~(Ref.~\cite{hastings2010locality}) }\\
  \hline
    \multirow{3}{*}{2D TFIM}   & $2X_0\sqrt{2Jh}$ & $ 4.27\sqrt{Jh}$  & \multirow{3}{*}{$16 e J$} &\multirow{3}{*}{$ 43.5 J$} \\
    \cline{2-3}
        & $8X_{\frac{1}{2}}J$      & $ 15.1 J$   &                                  &                          \\
    \cline{2-3}
                             &  $8X_0h $        & $ 12.1 h$   &                                  &                           \\ 
\hline
    \multirow{3}{*}{1D FH}   & $2 X_{\frac{3U}{4J}}J$ &  $4.14J_{(U=J)}$  &\multirow{3}{*}{$16 e J$}&\multirow{3}{*}{$ 43.5J$} \\
    \cline{2-3}
                              & $8X_0J$  &         $12.1 J$           &           &         \\
    \cline{2-3}
                             & $Z_{U/J}J$  &        $7.05J_{(U=5J)}$             &          &          \\
    \hline
      \multirow{3}{*}{PTC}           & $8X_{\frac{1}{2}}h$      & $ 15.1 h$   &     \multirow{3}{*}{$32 e $}    &  \multirow{3}{*}{$87.0$}\\
    \cline{2-3}
    & $2X_0\sqrt{2h}$ & $ 4.27\sqrt{h}$  &  & \\

    \cline{2-3}
                             &  $8X_0 $        & $ 12.1$   &                                  &           \\ 
    \hline
\end{tabular}}
\caption{\label{tab:comp} Comparison between previous LR speeds and those introduced in this paper, in the case of 2D spin-1/2 TFIM, 1D SU(2) FH model, and perturbed toric code~(PTC) model~\cite{kitaev2003fault} with on site $h\sum_j\hat{\sigma}^x_j$ perturbation. The LR speed $v_{\mathrm{LR}}$ of this paper is upper bounded by the minimum of all the expressions in the different rows of each model. For each column divided into two subcolumns, the first subcolumn shows analytic expressions,  and the second subcolumn shows their approximate numerical values. (For those bounds of the FH model depending on $U/J$ we have chosen a representative point).  The constant $X_y$ is defined as the solution to the equation $x\arcsinh(x)=\sqrt{x^2+1}+y$. $Z_{U/J}$ is defined in Eq.~\eqref{eq:vLRFHWy} and plotted in Fig.~\ref{fig:FHLRcomparison}.}
\end{table}

 \begin{table}[ht]
  \centering
  {\renewcommand{\arraystretch}{1.5}
  \begin{tabular}{|c|c|c|}
    \hline
    model   & $v_{\text{LR}}$~(this paper)  & $v_{\text{LR}}$~(Ref.\cite{hastings2010locality}) \\
    \hline
    large-$d$ TFIM    &             $\propto\sqrt{d}$                &         $\propto d$                   \\
    \hline
    large-$S$~(spin) TFIM &     $\propto\sqrt{S}$                    &         $\propto S^2$ \\
    \hline
   large-$N$ SU($N$) FH &           $\propto\sqrt{N}$               &  $\propto N$        \\
    \hline
Spin-$S$ Heisenberg  & const. & $\propto S$ \\
    \hline
    Wen's rotor model,    & \multirow{2}{*}{$\propto \sqrt{dgJ}$} & $\propto \sqrt{Jg} dS$~(Ref.\cite{Hamma2009PRL}) \\
     $S\to\infty$ and $J\ll g$       &  & $\propto d g$~(Ref.\cite{hastings2010locality}) \\
    \hline
   $N$-state truncated BH & $\propto\sqrt{N}$ & $\propto N$\\
    \hline
\end{tabular}}
\caption{\label{tab:compqual} Summary of qualitative improvements of $v_{\mathrm{LR}}$ in this paper for models with large on-site Hilbert spaces or large spatial dimensions: $S$ indicates the size of spins in spin models, $N$ the maximal number of particles on a single-site in Hubbard models, and $d$ the spatial dimension. In Wen's rotor model, we only list the result in the most interesting limit $S\to\infty,J\ll g$, where the previous best methods from Refs.~\cite{hastings2010locality,Hamma2009PRL} are infinitely looser than our result.}
\end{table}

The tighter LR bounds also improve several important results that rely on it. We mention two here. The first, which we explicitly derive in this paper, is that our LR bounds give rise to a tighter bound for the ground state correlation length in systems with a spectral gap. For example, when applied to the 1D TFIM, our bound gives $\xi\propto (\ln\frac{J}{h})^{-1}$ in the limit $J/h\to\infty$, in agreement with the exact solution, while the previous best bound approaches a constant; in the spin-$S$ Heisenberg model, when $S\to\infty$, our bound gives $\xi\leq c_1$~(a constant),  while the previous bound $\xi\leq c_2 (S+1)$ diverges linearly in $S$. As a second example, the improvement of the bound also enables new applications where quantitative accuracy is the key to get helpful results, for example in upper bounding the error of several numerical algorithms~\cite{Osborne2006,Osborne2007a,Osborne2007b,woods2016dynamical,tran2019locality}. Previous LR bounds would give extremely loose numerical error bounds, which renders the error bound practically useless in typical  situations where only modest system sizes can be simulated with current computational capabilities. We expect the tighter LR bounds to give practically useful numerical error bounds for reasonably small system sizes.

Our paper is organized as follows. In Sec.~\ref{sect:generalmethod} we outline the general method to upper bound the unequal time commutator by the solution of a set of first order linear differential equations on the commutativity graph. In Sec.~\ref{sec:TI} we specialize to translation invariant systems, where one can obtain Fourier integral solutions to the differential equations and obtain the bound in Eq.~\eqref{eq:fourierintegralsolu_intro}. We also derive an explicit formula for the LR speed by studying the analytic properties of the Fourier integrals. In Sec.~\ref{sec:factbound} we derive a power series solution to the aforementioned differential equations in an arbitrary graph, from which we prove the LR bound in Eq.~\eqref{eq:factorialLR}.
In Sec.~\ref{sec:examples} we give examples of our LR bounds in some paradigmatic models~(TFIM, spin-$S$ Heisenberg, $N$-state truncated BH, SU($N$) FH, and Wen's quantum rotor model).
In Sec.~\ref{sect:largeparam} we introduce a special treatment that can be used to obtain a tighter bound when some of the parameters in the Hamiltonian become large. In Sec.~\ref{sec:appl} we show how our LR bounds give a better bound on the ground state correlation length in systems with a spectral gap. In Sec.~\ref{sect:summary} we summarize our methods and results, and discuss their potential applications along with some future directions. In Appen.~\ref{appen:unifiedapproach} we briefly introduce a more general method that unifies some of the seemingly different approaches in the main text.

\section{General method: information propagation on commutativity graphs}\label{sect:generalmethod}
There are two major reasons why previous LR bounds are loose. First, the previous methods do not make use of the details of the Hamiltonian. For example, the derivations in Ref.~\cite{hastings2006} only makes use of how the operator norms of the interaction terms decay with interaction range, completely ignoring all other properties such as commutativity. Therefore, the previous results actually bound the ``worst case'' Hamiltonian that satisfies the finite range condition. Second, previous derivations use the integral inequalities iteratively, leading to a sum of products of terms along different paths. Since this sum over paths is difficult to calculate analytically, one has to use the triangle inequality many times to get a simple analytic expression, which makes the resulting bound looser.

In the following, we derive our LR bounds using a method that avoids these two difficulties. Motivated by the LR bound of a special model studied in Ref.~\cite{Hamma2009PRL} in the context of topological matter, we make use of the details of the Hamiltonian by carefully exploiting the commutation relations of the terms in the Hamiltonian. To facilitate this procedure, we introduce a graphical tool which we call the \textit{commutativity graph} that helps visualize the commutation relations of Hamiltonian terms, and will be introduced in Sec.~\ref{sect:definecommu}. In Sec.~\ref{sec:upperboundHeisengerg} we will use Heisenberg equation and triangle inequality to upper bound the unequal time commutator and will arrive at a similar integral inequality as previous LR bounds. However, instead of using it iteratively, we will find a set of differential equations whose solution naturally provides an upper bound for any quantities satisfying the integral inequalities. A simple analytic bound for solutions to these differential equations will be discussed in the next section, using a simpler method than the typical sum over paths in the previous approaches, and which produces much tighter bounds.

\subsection{The commutativity graph}\label{sect:definecommu}
We begin by introducing a useful graphical tool, namely the commutativity graph, to represent the Hamiltonian of locally interacting systems. Consider a locally-interacting quantum system in arbitrary dimension with an arbitrary lattice structure, or on an arbitrary graph. The Hamiltonian $\hat{H}$ can in general be written as
\begin{equation}\label{eq:Hgeneral}
\hat{H}=\sum_{j}h_j\hat{\gamma}_j,
\end{equation}
where $\hat{\gamma}_j$ are local Hermitian operators with unit norm $\|\hat{\gamma}_j\|=1$, and $h_j$ are constant parameters.

\begin{figure}
  \center{\includegraphics[width=\linewidth]{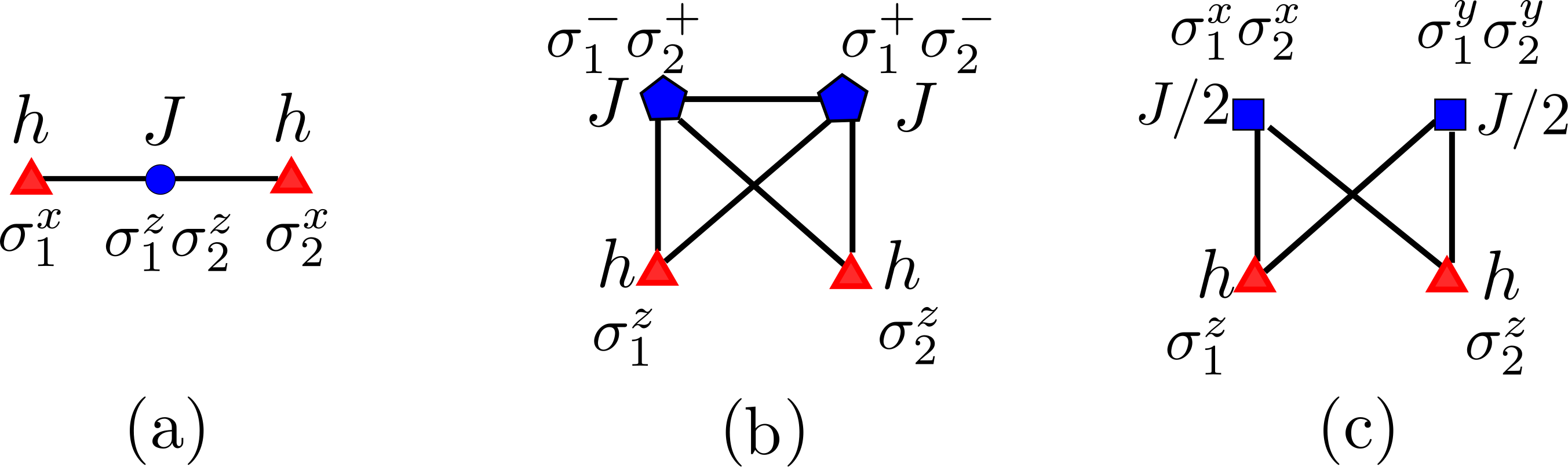}}
\caption{\label{fig:commuexample} Examples of commutativity graphs. (a) Commutativity graph of $\hat{H}_1=J\hat{\sigma}^z_1\hat{\sigma}^z_2+h(\hat{\sigma}^x_1+\hat{\sigma}^x_2)$. (b) Commutativity graph of $\hat{H}_2=J(\hat{\sigma}^-_1\hat{\sigma}^+_2+\hat{\sigma}^+_1\hat{\sigma}^-_2)+h(\hat{\sigma}^z_1+\hat{\sigma}^z_2)$. (c) Commutativity graph of $\hat{H}_2=\frac{J}{2}(\hat{\sigma}^x_1\hat{\sigma}^x_2+\hat{\sigma}^y_1\hat{\sigma}^y_2)+h(\hat{\sigma}^z_1+\hat{\sigma}^z_2)$.}
\end{figure}

The commutativity graph $G$ of the Hamiltonian $\hat{H}$ is constructed as follows. Each local operator $\hat{\gamma}_i$ is represented by a vertex $i$, the parameter $h_i$ is attached to vertex $i$,  and we link two vertices $i,j$ by an edge $\langle i,j \rangle$ if and only if the corresponding terms do not commute $[\hat{\gamma}_i,\hat{\gamma}_j]\neq 0$.  The resulting graph necessarily reflects the locality, since local operators acting on non-overlapping spatial regions must commute, so there is no edge between their representative vertices. Some simple examples of commutativity graphs are shown in Fig.~\ref{fig:commuexample}. Notice that the same Hamiltonian may have different decompositions in the form of Eq.~\eqref{eq:Hgeneral} and therefore different commutativity graphs, due to the freedom in how to partition terms of $\hat{H}$. Nevertheless, for convenience we will simply speak of ``the commutativity graph of $\hat{H}$'' when a certain decomposition is implicitly assumed. We will discuss how to choose the decomposition at the end of Sec.~\ref{sec:upperboundHeisengerg}. In the following we will derive an LR bound based on a differential equation on the commmutivity graph of $\hat{H}$.

\subsection{Upper bound for $\|[\hat{\gamma}_i(t),\hat{B}(0)]\|$}\label{sec:upperboundHeisengerg}
The Heisenberg equation for the operator $\hat{\gamma}_{i}(t)=e^{i\hat{H} t}\hat{\gamma}_{i}e^{-i \hat{H} t}$ is
\begin{equation}\label{eq:gammaHei}
i\dot{\hat{\gamma}}_{i}(t)=[\hat{\gamma}_{i}(t),\hat{H}]=\sum_{j:\langle ij\rangle\in G}h_j[\hat{\gamma}_{i}(t),\hat{\gamma}_{j}(t)],
\end{equation}
where the dot on $\hat{\gamma}_{i}(t)$ means time derivative, and the graph geometry provides a natural way to label the summation.

The general task of this paper is to find an upper bound for unequal time commutators $[\hat{A}(t),\hat{B}(0)]$ of local operators $\hat{A},\hat{B}$, \textit{i.e.} operators with finite support. Let us first consider the special case when $\hat{A}=\hat{\gamma}_{i}$ is a term of the Hamiltonian. In this case, we are interested in the unequal time commutator $\hat{\gamma}^B_{i}(t)=[\hat{\gamma}_{i}(t),\hat{B}(0)]$, for an arbitrary local operator $\hat{B}$. Using Eq.~\eqref{eq:gammaHei} and the Jacobi identity, we have the evolution equation for $\hat{\gamma}^B_{i}(t)$
\begin{equation}\label{eq:gammaBt}
  i\dot{\hat{\gamma}}^B_{i}(t)=\sum_{j:\langle ij\rangle\in G}h_j\{[\hat{\gamma}^B_{i}(t),\hat{\gamma}_{j}(t)]+
[\hat{\gamma}_{i}(t),[\hat{\gamma}_{j}(t),\hat{B}(0)]]\}.
\end{equation}
Substituting $\hat{\gamma}^B_{i}(t)=\hat{U}(t)\hat{\tau}^B_{i}(t)\hat{U}^\dagger(t)$ where $\hat{U}(t)$ is the unitary operator satisfying $\hat{U}(0)=1$ and $i\dot{\hat{U}}(t)=-\sum_{j:\langle ij\rangle\in G}h_j\hat{\gamma}_{j}(t)\hat{U}(t)$ into Eq.~\eqref{eq:gammaBt}, we have
\begin{eqnarray}\label{eq:GammaBt2}
  i\dot{\hat{\tau}}^B_{i}(t)= \hat{U}^\dagger(t)\sum_{j:\langle ij\rangle\in G}h_j[\hat{\gamma}_{i}(t),[\hat{\gamma}_{j}(t),\hat{B}(0)]]\hat{U}(t).
\end{eqnarray}
Now we bound the time dependence of the operator norm of ${\hat \gamma}_i^B(t)$, the fundamental object controlled by LR bounds. Since ${\hat \gamma}_i^B$ and ${\hat \tau}_i^B$ are related by unitary transforms, $\|\hat{\gamma}^B_{i}(t)\|=\|\hat{\tau}^B_{i}(t)\|$~(recall that $\|\hat{U}\hat{A}\hat{V}^\dagger\|=\|\hat{A}\|$ for unitary operators $\hat{U},\hat{V}$). Using this and applying the basic inequalities $\|\hat{A}+\hat{B}\|\leq \|\hat{A}\|+\|\hat{B}\|$ and $\|\hat{A}\cdot \hat{B}\|\leq \|\hat{A}\|\cdot\|\hat{B}\|$, we obtain
\begin{align}\label{ineq:gammaBt}
 \|\hat{\gamma}^B_{i}(t)\| -\|\hat{\gamma}^B_{i}(0)\| &=\|\hat{\tau}^B_{i}(t)\| -\|\hat{\tau}^B_{i}(0)\|\nonumber\\
 &\hspace{-0.35in} \leq\int^t_0\|\dot{\hat{\tau}}^B_{i}(t')\|dt'\nonumber\\
 &\hspace{-0.35in} \leq \sum_{j:\langle ij\rangle\in G}|h_j| \int^t_0\|[[\hat{\gamma}_{j}(t'),\hat{B}(0)],\hat{\gamma}_{i}(t')]\|dt'\nonumber\\
 & \hspace{-0.35in} \leq 2\sum_{j:\langle ij\rangle\in G} |h_j| \int^t_0 \|\hat{\gamma}^B_{j}(t')\|dt'.
\end{align}
where in the last line we used $\|\hat{\gamma}_{i}(t')\|=\|\hat{\gamma}_{i}\|=1$.

Although Eq.~\eqref{ineq:gammaBt} bounds the operator norm that is of interest, the right hand side depends on this unknown operator norm. Nevertheless, one can use the generalized Gr\"{o}nwall's inequality~\cite{Gronwall1976} to show that $\|\hat{\gamma}^B_{i}(t)\|\leq\bar{\gamma}^B_{i}(t)$, where ${\bar \gamma}_i^B(t)$ is the solution to Eq.~\eqref{ineq:gammaBt} with the inequality replaced by equality:
\begin{eqnarray}\label{eq:gammabarBt}
 \bar{\gamma}^B_{i}(t) =\bar{\gamma}^B_{i}(0) + 2\sum_{j:\langle ij\rangle\in G} |h_j| \int^t_0 \bar{\gamma}^B_{j}(t')dt',
\end{eqnarray}
where $\bar{\gamma}^B_{i}(0)=\|\hat{\gamma}^B_{i}(0)\|$. Differentiating Eq.~\eqref{eq:gammabarBt}, one obtains the first order linear differential equation
\begin{eqnarray}\label{eq:dfbart}
  \dot{\bar{\gamma}}^B_{i}(t) &= &2\sum_{j:\langle ij\rangle\in G}|h_j|\bar{\gamma}^B_j(t)
\end{eqnarray}
with initial condition
\begin{equation}\label{eq:initialcond}
  \bar{\gamma}^B_{i}(0)=\|[\hat{\gamma}_{i},\hat{B}]\|.
\end{equation}
We define the support of operator $\hat{B}$ on the commutativity graph as $S(\hat{B})=\{i|~[\hat{\gamma}_{i},\hat{B}]\neq 0\}$. Then the initial condition can be simplified as $\bar{\gamma}^B_{i}(0)=2\|\hat{B}\|(i\in S(\hat{B}))$, where the notation $(P)$ is defined as $(P)=1$ if $P$ is true and $(P)=0$ if $P$ is false, for an arbitrary statement $P$.

The differential equation Eq.~\eqref{eq:dfbart} gives an upper bound for the unequal time commutator of a Hamiltonian term with an arbitrary local operator.
To obtain the upper bound of $\hat{A}^B(t)=[\hat{A}(t),\hat{B}(0)]$, where $\hat{A}$ is an operator not in the Hamiltonian, we can perform similar calculations in Eqs.~(\ref{eq:gammaHei}-\ref{ineq:gammaBt}) to get
\begin{equation}\label{ineq:Zt}
  \|\hat{A}^B(t)\| -\|\hat{A}^B(0)\| \leq  \int^t_0 \sum_{i\in S(\hat{A})}2\|\hat{A}\||h_i| \bar{\gamma}^B_{i}(t')dt'.
\end{equation}

For later convenience, we introduce the Green's function as a useful tool to discuss solutions to Eq.~\eqref{eq:dfbart}. First notice that Eq.~\eqref{eq:dfbart} can be rewritten in a symmetric form
\begin{eqnarray}\label{eq:dfbart_2}
  \dot{\bar{\Gamma}}^B_{i}(t) &= &2\sum_{j:\langle ij\rangle\in G}\sqrt{|h_i| |h_j|}\bar{\Gamma}^B_j(t),
\end{eqnarray}
where $\bar{\Gamma}^B_{i}(t)=\sqrt{h_i}\bar{\gamma}^B_{i}(t)$. We denote the real symmetric coefficient matrix as $H_{ij}=2\sqrt{|h_i| |h_j|}(\langle ij\rangle\in G)$.
The Green's function $G_{ij}(t)$ is defined as the solution to the differential equation
\begin{equation}\label{eq:diffeqnGreen}
\dot{G}_{ij}(t)=\sum_{k:\langle ik\rangle\in G}H_{ik}G_{kj}(t),
\end{equation}
with initial condition $G_{ij}(0)=\delta_{ij}$, so any initial value problem can be obtained by linear combinations of the Green's functions:
\begin{equation}\label{eq:Greenbasicuse}
\bar{\Gamma}^B_{i}(t)=\sum_{j}G_{ij}(t)\bar{\Gamma}^B_{j}(0),
\end{equation}
and therefore the LR bound of $\|[\hat{\gamma}_i(t),\hat{B}_{Y}(0)]\|$ can be expressed as
\begin{eqnarray}\label{eq:gammaBintermsofgreen}
  \|[\hat{\gamma}_i(t),\hat{B}_{Y}(0)]\|\leq \sum_{j\in Y} G_{ij}(t)\sqrt{\frac{h_j}{h_i}}2\|\hat{B}\|,
\end{eqnarray}
where $Y=S(\hat{B})$ is the support of operator $\hat{B}$ on the commutativity graph~(from now on we use $\hat{A}_X$ to mean that operator $\hat{A}$ has support $X$ on the commutativity graph). The LR bound for the unequal time commutator between arbitrary local operators $\|[\hat{A}_X(t),\hat{B}_Y(0)]\|$ can be obtained by inserting Eq.~\eqref{eq:gammaBintermsofgreen} into Eq.~\eqref{ineq:Zt}:
\begin{align}\label{ineq:AXBYt0}
  \|[\hat{A}_X(t),\hat{B}_Y(0)]\| -\|[\hat{A}_X(0),\hat{B}_Y(0)]\|&\\
  &\hspace{-1.35 in}\leq  4\|\hat{A}\|\|\hat{B}\|\int^t_0 \sum_{i\in X,j\in Y} \sqrt{|h_ih_j|}G_{ij}(t')dt'.\nonumber
\end{align}

Eqs.~\eqref{eq:diffeqnGreen} and \eqref{ineq:AXBYt0}~[or Eqs.~\eqref{eq:dfbart} and \eqref{ineq:Zt}] are the main results of this section, which upper bound the unequal time commutators by the solution to a set of linear differential equations. Note that the bound Eq.~\eqref{ineq:AXBYt0} can in general be efficiently computed. One can obtain $G_{ij}(t)$ by solving Eq.~\eqref{eq:diffeqnGreen}, a set of $N_H$ linear coupled ordinary differential equations, where $N_H$ is the number of terms in the Hamiltonian, which grows linearly with system size in locally-interacting systems. This is a substantial reduction from solving the original many-body problem, whose cost in general grows exponentially with the system size. Furthermore, in translation invariant systems, we can solve these equations by a Fourier transform, which allows us to derive further simplified analytic upper bounds, as shown in the next section.

We end this section by commenting on the issue noted at the end of Sec.~\ref{sect:definecommu}. As we mentioned there, we can write the same Hamiltonian in the form of Eq.~\eqref{eq:Hgeneral} in different ways, leading to different commutativity graphs. The differential equations and the resulting LR bound will be different as well. One may ask how to choose a decomposition of the Hamiltonian $\hat{H}$ so that the resulting LR bound is tightest. While there is not a general statement about which decomposition is best, for spin-1/2 and fermionic systems we know a particularly good choice of decomposition that in many cases give us the tightest bound. For such systems, it is always possible to write the Hamiltonian in the form of Eq.~\eqref{eq:Hgeneral}~\cite{Nussinov2009bond,Cobanera2011bond} if we require that $\hat{\gamma}^2_i=1$ and $\hat{\gamma}_i,\hat{\gamma}_j$ for $i\neq j$ either commute or anti-commute. More precisely, the algebraic relations of $\{\hat{\gamma}_i\}$ are related to their commutativity graph $G$ as follows:
\begin{eqnarray}\label{eq:opalg}
  \hat{\gamma}^2_i&=&1,\text{ for } \forall i\in G,\nonumber\\
  \{\hat{\gamma}_i,\hat{\gamma}_{j}\}&=&0 \text{ for } \langle ij\rangle\in G,\nonumber\\
  \left[\hat{\gamma}_i,\hat{\gamma}_j\right]&=&0 \text{ for } \langle ij\rangle\notin G.
\end{eqnarray}
For example, in spin-$1/2$ systems $\hat{\gamma}_i$ can be chosen as products of Pauli matrices, while in fermionic systems they are products of Majorana fermion operators. More specific examples are given in Sec.~\ref{sec:examples}. If the additional constraints  Eq.~\eqref{eq:opalg} still fail to specify the decomposition uniquely, we would choose~[among those satisfying Eq.~\eqref{eq:opalg}] a decomposition such that the total number of $\hat{\gamma}_i$ terms is minimal. We call it a \textit{minimal Clifford decomposition}. The LR bound derived from this decomposition is typically tightest since the triangle inequality $\|[A,B]\|\leq 2\|A\|\|B\|$~(which is used many times in our derivation) is saturated when $\{A,B\}=0$. One can also show that the resulting LR bound has the tightest small-$t$ exponent in typical models, as discussed in Sec.~\ref{sec:factbound}.


\subsection{An alternative treatment for interacting fermions}\label{sect:fermions}
Our previous analysis applies to the fermionic case as well, since our derivations did not rely on any special property of the Hamiltonians beyond being locally-interacting. However, for the case of interacting fermions in an arbitrary $d$-dimensional lattice, we can use a slightly different method which sometimes gives us a tighter LR  bound. 

In the following we consider an arbitrary locally-interacting fermionic system with up to quartic terms, but the method generalizes straightforwardly to systems with higher order interacting terms. We will use the Majorana representation for convenience, where the Hamiltonian can in general be written as
\begin{equation}\label{eq:HMajgeneral}
\hat{H}=\sum_{i<j}t_{ij}i \hat{c}_{i}\hat{c}_{j}+\sum_{i<j<k<l}U_{ijkl}\hat{c}_{i}\hat{c}_{j}\hat{c}_{k}\hat{c}_{l}.
\end{equation}
where $t_{ij}$ and $U_{ijkl}$ are totally antisymmetric, and the Majorana operators satisfy
\begin{eqnarray}\label{eq:commuMajorana}
  \hat{c}_{i}^\dagger&=&\hat{c}_{i},\nonumber\\
  \{\hat{c}_{i},\hat{c}_{j}\}&=&2\delta_{ij}.
\end{eqnarray}

We first upper bound the norm of the unequal time anticommutator $\hat{c}_{ij}(t)=\{\hat{c}_{i}(t),\hat{c}_{j}(0)\}$. The Heisenberg equation for the operator $\hat{c}_{i}(t)=e^{i\hat{H} t}\hat{c}_{i}e^{-i \hat{H} t}$ is
\begin{equation}\label{eq:dcFHt}
  i\dot{\hat{c}}_{i}(t)=2\sum_{j}t_{ij} i \hat{c}_{j}(t)+2\sum_{j<k<l}U_{ijkl}\hat{c}_{j}(t)\hat{c}_k(t)\hat{c}_l(t).
\end{equation}
Using Eq.~\eqref{eq:dcFHt}, we have the evolution equation for $\hat{c}_{im}(t)$
\begin{eqnarray}\label{eq:dcbetatgeneral}
  i\dot{\hat{c}}_{im}(t)&=&   2\sum_{j<k<l}U_{ijkl}\{\hat{c}_{j}(t)\hat{c}_k(t)\hat{c}_l(t),\hat{c}_{m}(0)\}\nonumber\\
  &&{}+2\sum_{j} t_{ij}i \hat{c}_{jm}(t).
\end{eqnarray}
The first term of Eq.~\eqref{eq:dcbetatgeneral} can be expanded into the sum of three terms using $\{abc,d\}=ab\{c,d\}+a\{b,d\}c+\{a,d\}bc$.
Therefore
\begin{align}\label{ineq:dcbetat}
 \|\hat{c}_{im}(t)\| -\|\hat{c}_{im}(0)\|&\leq\int^t_0\|\dot{\hat{c}}_{im}(t')\|dt'\nonumber\\
  &\hspace{-.8in}\leq 2\int^t_0\sum_{j<k<l}|U_{ijkl}|\|\hat{c}_{jm}(t')+\hat{c}_{km}(t')+\hat{c}_{lm}(t')\|dt' \nonumber\\
         &\hspace{-.65in}{}+2\int^t_0\sum_{j} |t_{ij}|\| \hat{c}_{jm}(t')\|dt'.
\end{align}
Eq.~\eqref{ineq:dcbetat} is the analog of Eq.~\eqref{ineq:gammaBt}. Therefore, using the same argument as we used in Sec.~\ref{sec:upperboundHeisengerg}, we can conclude that $\|\hat{c}_{im}(t)\|\leq c_{im}(t)$ where $c_{im}(t)$ is the solution to the differential equation
\begin{eqnarray}\label{eq:dcbetat}
  \dot{c}_{im}(t)&=&2\sum_{j<k<l}|U_{ijkl}|[c_{jm}(t)+c_{km}(t)+c_{lm}(t)]\nonumber\\
  &&{}+2\sum_{j} |t_{ij}|c_{jm}(t),
\end{eqnarray}
with initial condition $c_{im}(0)=\|\hat{c}_{im}(0)\|=2\delta_{im}$. Once we know the LR bound for $\{\hat{c}_{i}(t),\hat{c}_{j}(0)\}$, we can calculate LR bounds for arbitrary local operators using identities like $[ab,cd]=a\{b,c\}d-ac\{b,d\}+\{a,c\}db-c\{a,d\}b$, since an arbitrary local operator can be expressed as products of basic Majorana operators $c_j$.

Eq.~\eqref{eq:dcbetat} takes a similar form as Eq.~\eqref{eq:dfbart} and can be treated similarly. For example, the Green's function method in Sec.~\ref{sec:upperboundHeisengerg} can also be applied, and the analyses in the following two sections apply to both equations equally well. We currently do not have a general criteria to determine which equation will lead to a tighter LR bound for a specific fermionic system, and the results have to be compared case by case. But compared to LR bounds obtained prior to the current paper, both methods give improved LR speeds, correct power law growth of correlations at short time, and the superexponentially decaying tail. See the application to the FH model in Sec.~\ref{sec:example_FH} for a detailed comparison. 

\section{Analytic bounds in translation invariant systems}\label{sec:TI}
In the last section we upper bounded the unequal time commutator by the solution to an integral inequality [Eq.~\eqref{ineq:AXBYt0}] with the integrand given by the solution to a set of linear differential equations [Eq.~\eqref{eq:diffeqnGreen}]. In translation invariant systems, we can obtain the solution to these equations in the form of a Fourier integral, and derive a simpler, fully analytic upper  bound by methods of complex analysis.

We will start our discussion  by considering Eq.~\eqref{eq:diffeqnGreen} in a general periodic lattice structure and obtain a formal solution in terms of a Fourier integral. Suppose there are $l$ vertices in a primitive unit cell, labeled by index $\alpha=1,\ldots,l$. We use capital $I$ to label unit cells with coordinate labeled by $\vec{r}_I$~[for example, if $\{\boldsymbol{\alpha}_1,\ldots,\boldsymbol{\alpha}_d\}$ is a set of primitive lattice translation vectors, then coordinate $\vec{r}_I=(n_1,n_2,\ldots, n_d),~n_a\in \mathbb{Z},~a=1,\ldots,d$ labels the primitive cell at physical position $\mathbf{R}_I=\sum^d_{a=1}n_a\boldsymbol{\alpha}_a$], so that every vertex $i$ can be labeled by a pair $(I,\alpha)$. Taking advantage of translation invariance, we can use the Fourier integral representation of the Green's function $G_{ij}(t)\equiv G_{I\alpha;J\beta}(t)$, 
\begin{equation}\label{eq:Greenrt}
G_{ij}(t)\equiv \int^\pi_{-\pi}\frac{d^dk}{(2\pi)^d}  G^{(\vec{k})}_{\alpha\beta}(t) e^{i\vec{k}\cdot(\vec{r}_I-\vec{r}_J)}.  
\end{equation}
[Notice that other discrete symmetries of the lattice~(such as inversion, reflection, discrete rotation, etc.) do not play a significant role in this context, so we are using integer coordinates of unit cells and consequently the Fourier momentum simply lies in $(-\pi,\pi]^d$ rather than the first Brillouin zone in typical  solid state contexts.] Inserting Eq.~\eqref{eq:Greenrt} into Eq.~\eqref{eq:diffeqnGreen} we get
\begin{eqnarray}\label{eq:dfbart_k}
  \dot{G}^{(\vec{k})}_{\alpha\beta}(t)=\sum^l_{\gamma=1}H^{(\vec{k})}_{\alpha\gamma}G^{(\vec{k})}_{\gamma\beta}(t),
\end{eqnarray}
where the $l\times l $ coefficient matrix $H^{(\vec{k})}_{\alpha\beta}$ is the Fourier transform of $H_{I\alpha;J\beta}$~(notice that due to translation invariance $H_{I\alpha;J\beta}$ only depends on $\vec{r}_I-\vec{r}_J$):
\begin{equation}\label{eq:expansionHk}
H^{(\vec{k})}_{\alpha\beta}=\sum_J H_{I\alpha;J\beta}e^{-i\vec{k}\cdot(\vec{r}_I-\vec{r}_J)}.
\end{equation}
The formal solution to Eq.~\eqref{eq:dfbart_k} is $G^{(\vec{k})}_{\alpha\beta}(t)=[e^{H^{(\vec{k})}t}]_{\alpha\beta}$, where $H^{(\vec{k})}$ is considered as an $l\times l$ matrix.

\begin{figure}
  \center{\includegraphics[width=0.6\linewidth]{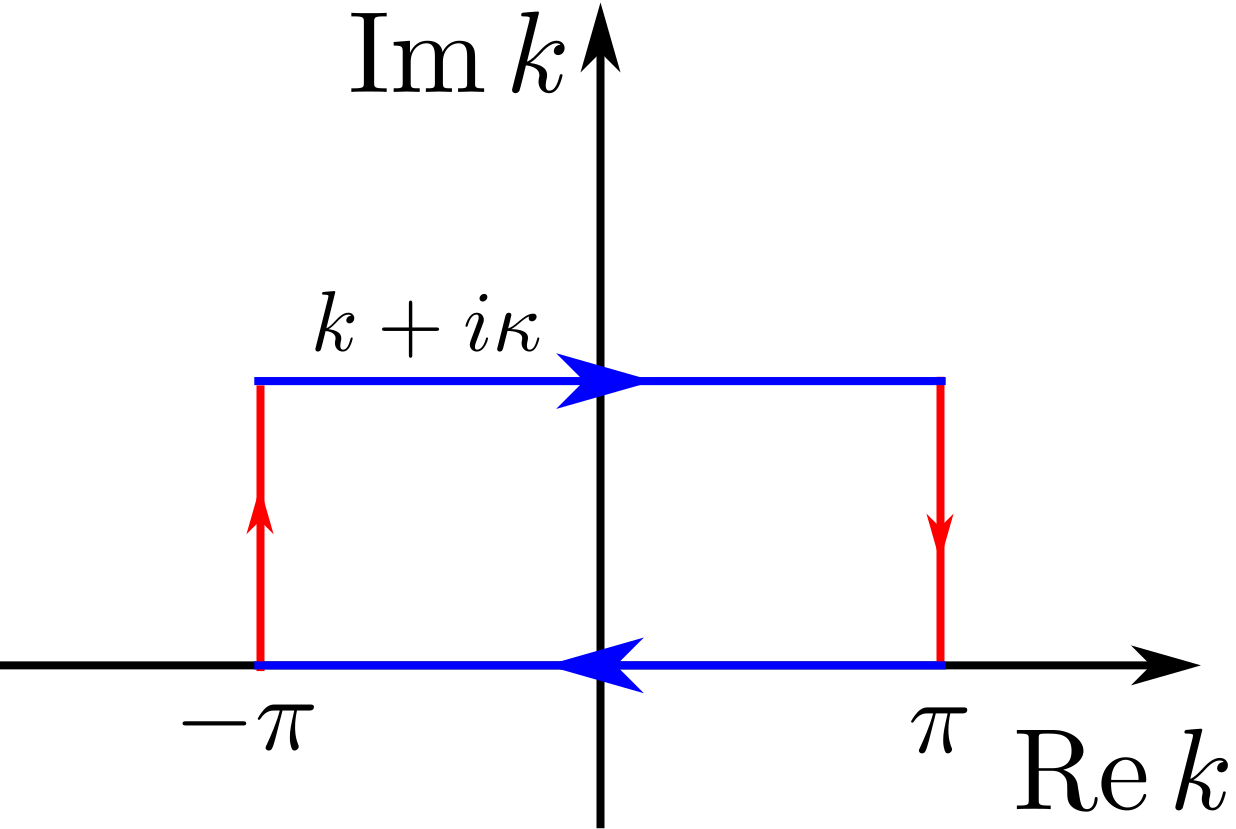}}
\caption{\label{fig:contourdeform}  Since the integrand of Eq.~\eqref{eq:Greenrt} is a complex analytic function of $\vec{k}$ except at infinity, the integration along the above contour is zero. Since the integrand is a periodic function of $\vec{k}$ with period $2\pi$, the integral on two vertical arrows~(colored red) cancel. The net result is that in Eq.~\eqref{eq:Greenrt} we can send $\vec{k}\to\vec{k}+i\vec{\kappa}$ without changing the result of the integral. }
\end{figure}

We now derive a simple analytic upper bound for the Fourier integral of Eq.~\eqref{eq:Greenrt} using methods of complex analysis. Since the integrand of Eq.~\eqref{eq:Greenrt} is a complex analytic function of $\vec{k}$ except at infinity, and since the integrand is periodic in shifts of the real axis by $2\pi$, we can send $\vec{k}\to\vec{k}+i\vec{\kappa}$ without changing the result of the integral, as illustrated in Fig.~\ref{fig:contourdeform}. Therefore we have
\begin{eqnarray}\label{eq:Greenrtbound}
  |G_{ij}(t)|&=& \left|\int^\pi_{-\pi}\frac{d^dk}{(2\pi)^d} [e^{H^{(\vec{k}+i\vec{\kappa})} t}]_{\alpha\beta} e^{i\vec{k}\cdot\vec{r}-\vec{\kappa}\cdot\vec{r}}\right|\nonumber\\
              &\leq& \int^\pi_{-\pi}\frac{d^dk}{(2\pi)^d} |[ e^{H^{(\vec{k}+i\vec{\kappa})} t}]_{\alpha\beta}| e^{-\vec{\kappa}\cdot\vec{r}}\nonumber\\
              &\leq& |[ e^{H^{(i\vec{\kappa})} t}]_{\alpha\beta}| e^{-\vec{\kappa}\cdot\vec{r}}\nonumber\\
              &\leq& c_{\vec{\kappa}}  e^{\omega_m(i\vec{\kappa})t-\vec{\kappa}\cdot\vec{r}},
\end{eqnarray}
where $\vec{r}=\vec{r}_I-\vec{r}_J$, in the second line we used the triangle inequality for the integral, and in the third line we used the inequality $|[ e^{H^{(\vec{k}+i\vec{\kappa})}t}]_{\alpha\beta}|\leq |[ e^{H^{(i\vec{\kappa})}t}]_{\alpha\beta}|$~[which follows from the  triangle inequality since the coefficients $H_{I\alpha;J\beta}$ in Eq.~\eqref{eq:expansionHk} are non-negative]. In the last line we used the inequality  $|[ e^{H^{(i\vec{\kappa})}t}]_{\alpha\beta}|\leq \|e^{H^{(i\vec{\kappa})}t}\|\leq c_{\vec{\kappa}}e^{\omega_m(i\vec{\kappa}) t}$, where $\omega_m(i\vec{\kappa})$ is the eigenvalue of $H^{(i\vec{\kappa})}$ with largest magnitude~[$\omega_m(i\vec{\kappa})$ must be real positive according to the Perron-Frobenius theorem], $c_{\vec{\kappa}}$ is a function of $\vec{\kappa}$ independent of $t$, and the last inequality can be proved by diagonalizing $H^{(i\vec{\kappa})}$ into Jordan canonical form. Therefore we arrive at the important result:
\begin{equation}\label{eq:upperboundgreen}
|G_{ij}(t)|\leq c_{\vec{\kappa}} e^{\omega_m(i\vec{\kappa})t-\vec{\kappa}\cdot\vec{r}}, ~~\forall \vec{\kappa}\in \mathbb{R}^d.
\end{equation}
While this expression looks quite similar to previous LR bounds with exponential decaying tail, described by Eq.~\eqref{eq:previousLR}, the point here is that we can choose $\vec{\kappa}=\vec{\kappa}(\vec{r},t)$ to depend on $\vec{r},t$ to minimize the RHS at each point in space-time. This makes the bound decaying faster than exponential at large distance, as we prove in Sec.~\ref{sec:factbound}, and see in examples in Sec.~\ref{sec:examples}.

We end this subsection by giving an explicit formula for the LR speed that emerges from this bound. The LR speed of the Green's function $G_{ij}(t)$~(considered as a function of $\vec{r},t$) in Eq.~\eqref{eq:Greenrt} is, by definition, the speed at which $r$ must change as a function of $t$ in order for $G_{ij}(t)$'s magnitude to stay constant. While the Eq.~\eqref{eq:Greenrt} is not simple enough to exactly extract the LR speed, we can get an upper bound for the LR speed from the upper bound of $G_{ij}(t)$ in Eq.~\eqref{eq:upperboundgreen}. If we choose $\vec{\kappa}=\vec{\kappa}_0\equiv\mathrm{sgn}(\vec{r})\kappa_0$~[where $\mathrm{sgn}(\vec{r})\equiv(\mathrm{sgn}(r_1),\mathrm{sgn}(r_2),\ldots,\mathrm{sgn}(r_d))$], we get the old form of LR bound with exponential decaying tail
\begin{equation}\label{eq:relationtooldLR}
  |G(\vec{r},t)|\leq c_{\vec{\kappa}_0} e^{\omega_m(i\vec{\kappa}_0)t-\kappa_0 |\vec{r}|},
\end{equation}
where $|\vec{r}|=\sum^d_{j=1}r_j$ is the Manhattan distance between the corresponding unit cells. Therefore the LR speed of $|G(\vec{r},t)|$ must be upper bounded by an optimal choice of $\kappa_0$:
\begin{equation}\label{eq:v_LR}
v_{\mathrm{LR}}\leq \min_{\kappa_0>0}\frac{\omega_m(i\vec{\kappa}_0)}{\kappa_0}.
\end{equation}
Notice that while $\vec{\kappa}_0$ depends on  $\mathrm{sgn}(\vec{r})$, in typical models $\omega_m(i\vec{\kappa}_0)$ does not depend on $\mathrm{sgn}(\vec{r})$~[see Sec.~\ref{sec:examples}]. In the case that $\omega_m(i\vec{\kappa}_0)$ does depend on $\mathrm{sgn}(\vec{r})$, Eq.~\eqref{eq:v_LR} would represent a directional LR speed, and to get the overall LR speed we should take the maximum over all possible $2^d$ values of $\mathrm{sgn}(\vec{r})$.

It is immediately clear from Eq.~\eqref{ineq:AXBYt0} that the LR speed of the Green's function gives an upper bound for the LR speed of the unequal time commutator between any local operators $\|[\hat{A}_{X}(t),\hat{B}_{Y}(0)]\|$, since, by inserting  Eq.~\eqref{eq:relationtooldLR} into Eq.~\eqref{ineq:AXBYt0}, we get
\begin{eqnarray}\label{eq:AXBYLRspeed}
  \|[\hat{A}_X(t),\hat{B}_Y(0)]\|  \leq C e^{\omega_m(i\vec{\kappa}_0)t-\kappa_0 r_{XY}},
\end{eqnarray}
where $C=4\|\hat{A}\|\|\hat{B}\|\sqrt{h_Xh_Y}c_{\vec{\kappa}_0}/\omega_m(i\vec{\kappa}_0)$, $h_X\equiv\max_{i\in X}h_i$, $r_{XY}=\min_{i\in X,j\in Y}r_{ij}$ and we assumed $X\cap Y=\emptyset$. This proves that $\|[\hat{A}_X(t),\hat{B}_Y(0)]\|$ has the same LR speed $v_{\text{LR}}$.

Equation~\eqref{eq:v_LR} is one of the main results of our paper, for upper bounding the LR speed of a locally interacting many body Hamiltonian. Compared to previous methods for upper bounding $v_{\text{LR}}$, for example the methods given in Ref.~\cite{hastings2010locality} or Ref.~\cite{bravyi2006}, our formula Eq.~\eqref{eq:v_LR} is not only simpler but also gives much tighter bounds, as will be shown in specific examples in Sec.~\ref{sec:examples}.
\section{A simple LR bound for an arbitrary graph}\label{sec:factbound}
In this section we will prove the simple LR bound in Eq.~\eqref{eq:factorialLR} for an arbitrary locally interacting system. The only assumption we make about the system is that the commutativity graph $G$ has an upper bound on the degree of vertices and the weight $H_{ij}$ of the edges. We will first express the Green's function in an arbitrary graph as a Taylor series of $t$, illustrate the graph-theoretical meaning of the expansion coefficients, and then use this expansion to prove the LR  bound.

The formal solution to Eq.~\eqref{eq:diffeqnGreen} is
\begin{equation}\label{eq:formalsoluGreen}
G_{ij}(t)=[e^{H t}]_{ij}=\sum_{n\geq 0}G^{(n)}_{ij}\frac{t^n}{n!},
\end{equation}
where $G^{(n)}_{ij}=[H^n]_{ij}$. Notice that $H_{ij}$ can be considered as the adjacency matrix of the weighted commutativity graph $G$ with the weight of the edge $(i,j)$ being $H_{ij}$. In this way, the meaning of $G^{(n)}_{ij}$  is the sum of the weights of all possible paths in $G$ connecting vertices $i$ and $j$, where the weight of a path is the product of the weights of all its edges. [Note: there can be duplicate edges in a path, and the weight of path $p$ is $w_p=\prod_{e\in p}w^{d_e}_e$ where $d_e$ is the multiplicity of edge $e$.] Therefore, we have $G^{(n)}_{ij}=0$ for $n<d_{ij}$ where $d_{ij}$ is the graph theoretical distance~(length of the shortest path) between $i$ and $j$. In summary we have
\begin{equation}\label{eq:formalsoluGreensum}
G_{ij}(t)=\sum_{n\geq d_{ij}}G^{(n)}_{ij}\frac{t^n}{n!}.
\end{equation}

We now prove the LR bound in Eq.~\eqref{eq:factorialLR}. 
First let us prove that there exists a velocity $u$ such that for all pairs of vertices $i,j$ we have
\begin{equation}\label{eq:greenfactorialbound}
G_{ij}(t)\leq c(ut/d_{ij})^{d_{ij}},\text{ for } t\leq d_{ij}/u,
\end{equation}
for some constant $c$. Inserting Eq.~\eqref{eq:formalsoluGreensum}, the above inequality is equivalent to
\begin{equation}\label{eq:Greenboundproof0}
\sum_{n\geq d_{ij}}G^{(n)}_{ij}\frac{t^{n-d_{ij}}}{n!}\leq c (u/d_{ij})^{d_{ij}}, \text{ for } t\leq d_{ij}/u.
\end{equation}
Since the $H_{ij}$ are non-negative, the coefficients $G^{(n)}_{ij}$ are non-negative, so the LHS of Eq.~\eqref{eq:Greenboundproof0} is a non-decreasing function of $t$ while the RHS is independent of $t$. Therefore we only need to prove that the inequality holds for $t=d_{ij}/u$:
\begin{equation}\label{eq:Greenboundproof2}
\sum_{n\geq d_{ij}}G^{(n)}_{ij}\frac{(d_{ij}/u)^{n-d_{ij}}}{ n!}\leq c (u/d_{ij})^{d_{ij}},
\end{equation}
which can be simplified to
\begin{equation}\label{eq:Greenfinitevelocitycond}
  G_{ij}(d_{ij}/u)\leq c.
\end{equation}
This is exactly the statement that the Green's function has a finite LR velocity $u$. 

One can show that Eq.~\eqref{eq:Greenfinitevelocitycond} holds extremely generally, requiring only locality and bounded magnitude of the $H_{ij}$. For translation invariant systems, we have already proved this in the last section~[see Eq.~\eqref{eq:relationtooldLR}], except that distance is measured differently in Eq.~\eqref{eq:relationtooldLR} and Eq.~\eqref{eq:Greenfinitevelocitycond}: in the former case we used the Manhattan distance between unit cells while in the latter case we use graph-theoretical distance in the commutativity graph, so $u$ is in general different from $v_{\text{LR}}$. But since $c_1\leq d_{ij}/r_{ij}\leq c_2$ for some constants $c_1,c_2$~(this follows straightforwardly from the fact that each real space unit cell is associated with a finite number of terms in $\hat{H}$), existence of a finite $v_{\text{LR}}$ guarantees existence of $u$. In the next section we will show in some specific models how to find a tight $u$. For the general case without translation invariance, existence of a finite $u$ can be proved under the assumption that the weighted graph $G$ has an upper bound on the degree of vertices and the weights  of the edges, that is, every vertex has at most $\lambda$ edges and the weights satisfy $H_{ij}\leq h$ for all pairs of neighboring vertices $i,j$. With these assumptions we have $G^{(n)}_{ij}\leq h^n \lambda^n,~\forall i,j\in G,~n\geq 0$. Inserting into Eq.~\eqref{eq:formalsoluGreensum}, we have
\begin{eqnarray}\label{eq:appenproof1}
  G_{ij}(t)&\leq& \sum_{n\geq d_{ij}}\frac{(\lambda h t)^n}{n!}\nonumber\\
           &\leq& \sum_{n\geq d_{ij}}\frac{(\lambda h t)^n}{e(n/e)^n}\nonumber\\
           &\leq& \sum_{n\geq d_{ij}}(\lambda h te/d_{ij})^n/e\nonumber\\
           &=& (\lambda h te/d_{ij})^{d_{ij}}\frac{1/e}{1-\lambda h te/d_{ij}},
\end{eqnarray}
for $t<d_{ij}/\lambda he $. This proves Eq.~\eqref{eq:Greenfinitevelocitycond} if we choose, for example, $u=2\lambda h e$.


Now we use the bound Eq.~\eqref{eq:greenfactorialbound} on $G_{ij}(t)$ to bound a general unequal-time commutator $\|[\hat{A}_X(t),\hat{B}_{Y}(0)]\|$. Inserting Eq.~\eqref{eq:greenfactorialbound} into Eq.~\eqref{ineq:AXBYt0}, we have
\begin{align}\label{eq:ABfactorialbound_deri}
  \hspace{-.13in}\|[\hat{A}_X(t),\hat{B}_{Y}(0)]\|-\|[\hat{A}_X(0),\hat{B}_{Y}(0)]\|&\nonumber\\
  &\hspace{-1.5in}\leq 4 c_{\vec{\kappa}_0}\|\hat{A}\|\|\hat{B}\|\int^t_0\sum_{i\in X,j\in Y}\sqrt{h_ih_j}\left(\frac{ut'}{d_{ij}}\right)^{d_{ij}}dt'\nonumber\\
  &\hspace{-1.5in}\leq 4c_{\vec{\kappa}_0}\|\hat{A}\|\|\hat{B}\|\frac{e}{u}\sum_{i\in X,j\in Y}\sqrt{h_ih_j}\left(\frac{ut}{d_{ij}+1}\right)^{d_{ij}+1}\nonumber\\
  &\hspace{-1.5in}\leq 4c_{\vec{\kappa}_0}\|\hat{A}\|\|\hat{B}\|\frac{e}{u}h_{XY}\left(\frac{ut}{d_{XY}+1}\right)^{d_{XY}+1}
\end{align}
for $t\leq d_{XY}/u$, where  $h_{XY}$ is a geometric factor defined as
\begin{equation}\label{eq:hXY}
h_{XY}=\sum_{i\in X,j\in Y} \sqrt{h_ih_j}e^{d_{XY}-d_{ij}}.
\end{equation}
In the third line of Eq.~\eqref{eq:ABfactorialbound_deri} we used the inequality$(1+1/d_{ij})^{d_{ij}}\leq e$, and in the last line  we used $(ut/x)^{x}\leq (ut/y)^{y}e^{y-x}$ for $t\leq y/u\leq x/u$ ~[this inequality holds since the function $(ut/x)^{x}e^x$ is decreasing in $x$ for $x\geq ut$]. For $t> d_{XY}/u$ we have the trivial bound $\|[\hat{A}_X(t),\hat{B}_{Y}(0)]\|\leq 2\|\hat{A}\|\|\hat{B}\|\leq 2e\|\hat{A}\|\|\hat{B}\|\left(\frac{ut}{d_{XY}+1}\right)^{d_{XY}+1}$. In summary we have
\begin{equation}\label{eq:ABfactorialbound}
  \|[\hat{A}_X(t),\hat{B}_Y(0)]\|  \leq  C \left(\frac{ut}{d_{XY}+1}\right)^{d_{XY}+1},
\end{equation}
where $C=2e\max\{1,2c_{\vec{\kappa}_0}\frac{h_{XY}}{u}\} \|A\|\|B\|$ for $X\cap Y=\emptyset$ and $\forall t\geq 0$. This completes the proof of the LR bound in Eq.~\eqref{eq:factorialLR}. [Eq.~\eqref{eq:ABfactorialbound} is actually a slightly tighter version].


The LR bound in Eq.~\eqref{eq:ABfactorialbound} substantially and qualitatively improves previous LR bounds, not only because it has the superexponential decaying tail $e^{-r\ln r}$, but also because it has much tighter small-$t$ exponent, and a much tighter LR speed, as will be seen in examples in the next section. The small-$t$ exponent $\eta(\vec{r})$ of a bound $f(\vec{r},t)$ is defined by its limiting behavior $f(\vec{r},t)\propto t^{\eta(\vec{r})}$ as $t\to 0$. We now argue that if a minimal Clifford decomposition~[as defined below Eq.~\eqref{eq:opalg}] of the Hamiltonian is used in Eq.~\eqref{eq:Hgeneral}, then the bound in Eq.~\eqref{eq:ABfactorialbound} has a small-$t$ exponent that generically agree with the exact result, and thus are the tightest possible. For simplicity, consider the LR bound for $[\hat{\gamma}_i(t),\hat{\gamma}_{j}(0)]$.
The Baker-Campbell-Hausdorff formula gives
\begin{eqnarray}\label{eq:gammaBCBHexpansion}
  [\hat{\gamma}_i(t),\hat{\gamma}_{j}]&=&\left[\sum_{n\geq 0}\frac{(it)^n}{n!}\mathrm{ad}^n_{\hat{H}}( \hat{\gamma}_i),\hat{\gamma}_{j}\right]\\
  &=&\frac{(2it)^{d_{ij}-1}}{(d_{ij}-1)!}\sum_{\substack{p\in P_{ij}}} \hat{\gamma}_i\hat{S}_p\hat{\gamma}_{j}+O(t^{d_{ij}}),\nonumber
\end{eqnarray}
where $\mathrm{ad}_{\hat{H}}$ is the adjoint map of $\hat{H}$ acting on the space of operators defined as $\mathrm{ad}_{\hat{H}}(\hat{B})\equiv [\hat{H},\hat{B}]$,  $P_{ij}$ is the set of shortest paths connecting $i$ and $j$~(note: by our convention a path $p\in P_{ij}$ does not contain $i$ and $j$), the string operator $\hat{S}_p$ along a path $p\in P_{ij}$ is defined as $\hat{S}_p=\prod_{k\in p}h_k\hat{\gamma}_k$~(with a suitable ordering). Therefore the small-$t$ exponent is $\eta_{ij}=d_{ij}-1$ provided that the sum on the second line of Eq.~\eqref{eq:gammaBCBHexpansion} is nonzero. A sufficient condition for this to be true is if no two string operators $\hat{S}_{p}$ and $\hat{S}_{p'}$ along different paths $p\neq p'$ are proportional to each other~[Notice that each individual term $\hat{\gamma}_i\hat{S}_p\hat{\gamma}_{j}$ is a product of invertible operators~(due to the relation $\hat{\gamma}^2_j=1$) and is therefore always nonzero]. This condition is satisfied by typical models such as the TFIM, FH model, the Heisenberg XYZ model, and we believe that it is automatically satisfied by the minimal Clifford decomposition of the Hamiltonian.  Therefore the small-$t$ exponent $d_{XY}+1$ of the bound in Eq.~\eqref{eq:ABfactorialbound} is saturated by exact results, since when $X=S(\hat{\gamma}_i),Y=S(\hat{\gamma}_j)$, we have $d_{ij}-1=d_{XY}+1$. This has never been achieved in previous LR bounds, where the small-$t$ exponent is typically just $\eta(\vec{r})=1$, as in Eq.~\eqref{eq:previousLR} for example~\footnote{Refs.~\cite{bravyi2006,Osbornesimple2006,Lucas2019operator,chessa2019time} also derived LR bounds which decay superexponentially in distance, and have small-$t$ exponents proportional to the Manhattan distance in real space. But these bounds are still looser than ours in most cases of TFIM. For example, their small-$t$ exponent is $|\vec{r}|$ while our bound in Eq.~\eqref{eq:LRboundIsingsimple1} has small-$t$ exponent $2|\vec{r}|$, and at large $r\equiv |\vec{r}|$, their bound decays like $e^{-r\ln r}$ while our bound decays like $e^{-2 r\ln r}$, due to the fact that the distance on the commutativity graph of TFIM is twice the distance on real space. Also, Ref.~\cite{Osbornesimple2006} only gives the 1D result and the method is hard to generalize to higher dimensions, while the higher dimensional cases of Refs.~\cite{bravyi2006,Lucas2019operator} requires solving a hard combinatorical problem of summing over paths, the computational difficulty of which may grow exponentially with system size.}. 

\section{Examples}\label{sec:examples}
In this section, we apply the techniques developed above to calculate the LR velocity for the bound of Eq.~\eqref{eq:factorialLR} explicitly for five models as examples: the $d$-dimensional TFIM~(Sec.~\ref{sect:Ising}), the spin-$S$ Heisenberg model~(Sec.~\ref{sec:Heisenberg-S}), the $N$-state truncated BH model~(Sec.~\ref{sec:truncBH}), the SU($N$) FH model~(Sec.~\ref{sec:example_FH}), and Wen's quantum rotor model~(Sec.~\ref{sec:Wenrotor}). The techniques are general and can be straightforwardly applied to arbitrary models. All these examples demonstrate significant quantitative improvements over the previous LR speeds, and in a number of limits~(e.g. large-$d$, large-$S$ or large-$N$) our bounds are qualitatively tighter for having better scalings. We will introduce techniques to further tighten the LR velocity in Sec.~\ref{sect:largeparam}.
\subsection{The $d$-dimensional TFIM}\label{sect:Ising}
In the following we will first derive a bound in the familiar spin-1/2 case, and then extend this result to arbitrary spin in Sec.~\ref{sec:Ising-S}. 
\subsubsection{Spin-1/2 case}\label{sec:Ising-1/2}
\begin{figure}
  \center{\includegraphics[width=0.6\linewidth]{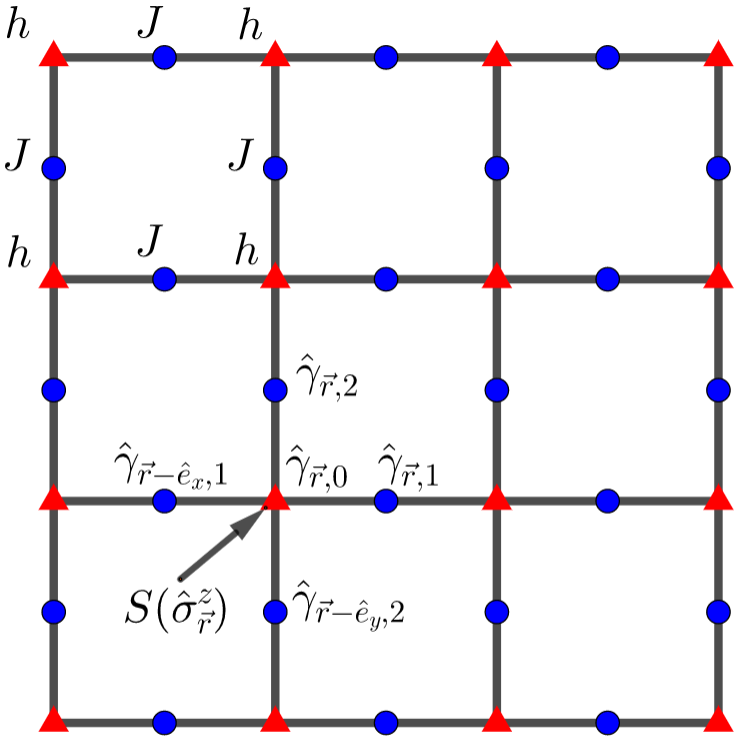}}
\caption{\label{fig:IsingCommu}  The commutativity graph $G$ of the 2D TFIM. Each term~(a local product of Pauli matrices) of the Hamiltonian is represented by a vertex. Any pair of terms either commutes or anticommutes. We link two vertices by an edge if the corresponding operators anticommute. The coefficient of each term is drawn adjacent  to the corresponding vertices. The arrow points to the region $S(\hat{\sigma}^z_{\vec{r}})$, the support of operator $\hat{\sigma}^z_{\vec{r}}$, which consists of only one point in this case.}
\end{figure}

The Hamiltonian for the $d$-dimensional hypercubic lattice Ising model with a transverse field is
\begin{equation}\label{eq:tIsingH}
\hat{H}=-J\sum_{\vec{r},1\leq j\leq d}\hat{\gamma}_{\vec{r},j} -h\sum_{\vec{r}}\hat{\gamma}_{\vec{r},0},
\end{equation}
with $\hat{\gamma}_{\vec{r},j}=\hat{\sigma}^z_{\vec{r}}\hat{\sigma}^z_{\vec{r}+\hat{e}_j}$ where $\hat{e}_j$ is the unit vector in the $j$\ts{th} direction, and $\hat{\gamma}_{\vec{r},0}=\hat{\sigma}^x_{\vec{r}}$. We will assume $J\ge 0$, $h\ge 0$~(the resulting LR bound only depends on $|J|$ and $|h|$). We have written the Hamiltonian in this form so that Eq.~\eqref{eq:opalg} is satisfied. It is easily verified that the square of each term is equal to unity $\hat{\gamma}^2_{\vec{r},\alpha}=1$, $\alpha=0,1,2,\ldots,d$, and any two terms either commute or anticommute. Fig.~\ref{fig:IsingCommu} shows the commutativity graph of this model for the $d=2$ case.

We can use the general Fourier integral representation of the Green's function in Eq.~\eqref{eq:Greenrt}
\begin{equation}\label{eq:Greenrt_Ising}
G_{\alpha\beta}(\vec{r},t)\equiv \int^\pi_{-\pi}\frac{d^dk}{(2\pi)^d}  [\exp(H^{(\vec{k})}t)]_{\alpha\beta} e^{i\vec{k}\cdot(\vec{r}_I-\vec{r}_J)},  
\end{equation}
where $H^{(\vec{k})}$, defined in Eq.~\eqref{eq:expansionHk}, is in this case a $(d+1)\times (d+1)$ matrix
\begin{equation}\label{eq:Hk_Ising}
  H^{(\vec{k})}=2\sqrt{Jh}\begin{bmatrix}
    0&1+e^{-i k_1}&\cdots &1+e^{-i k_d}\\
    1+e^{i k_1}&0&0&0\\
    \vdots&0&0&0\\
    1+e^{i k_d}&0&0&0\\
    \end{bmatrix},
  \end{equation}
so that
\begin{eqnarray}\label{eq:expHk_Ising}
  \exp[H^{(\vec{k})}t]&=&\frac{\sinh[\omega(\vec{k})t]}{\omega(\vec{k})}H^{(\vec{k})}+\cosh[\omega(\vec{k})t]\mathds{1}\nonumber\\
  &&+P_0\{1-\cosh[\omega(\vec{k})t]\},
\end{eqnarray}
where $\mathds{1}$ is the identity matrix and
\begin{eqnarray}\label{eq:P_0}
  P_0&=&\begin{bmatrix}0&0\\0& \mathds{1}-\Omega_{\vec{k}}\cdot\Omega_{\vec{k}}^\dagger\end{bmatrix},\nonumber\\
  \Omega_{\vec{k}}&=&\frac{1}{\sqrt{2d+2\cos\vec{k}}}
  \begin{bmatrix}
    1+e^{ik_1}\\
    \vdots\\
    1+e^{ik_d}
  \end{bmatrix},
  \end{eqnarray}
with $\cos\vec{k}\equiv \sum^d_{j=1}\cos k_j$. The eigenvalues of $H^{(\vec{k})}$ are $0$ and $\omega^{\pm}(\vec{k})=\pm\omega(\vec{k})=\pm\sqrt{8Jh(d+\cos\vec{k})}$, from which we can compute the upper bound for the LR speed given by Eq.~\eqref{eq:v_LR}:
\begin{equation}\label{eq:vLRIsing}
v_{\mathrm{Ising}}\leq 2X_0\sqrt{dJh},
\end{equation}
where the constant $X_y$ is defined as the solution to the equation $x\arcsinh(x)=\sqrt{x^2+1}+y$, and $X_0\approx 1.50888$. The  previous best bound for $v_{\mathrm{Ising}}$ is $8edJ\approx 21.7 dJ$ obtained from the method in Ref.~\cite{hastings2010locality}, so our bound in Eq.~\eqref{eq:vLRIsing} is a major, parametric improvement when $J$ or $d$ is large. Here we emphasize that to our knowledge this is the first LR bound whose speed scales sublinearly with spatial dimension $v\propto \sqrt{d}$--all previously known LR bounds have velocities that scale linearly with $d$~\footnote{The reason for this is that most of the previous LR bounds are very generic to include a large family of models. As we will see at the end of Sec.~\ref{sec:example_FH}, the speed of $d$-dimensional free fermions grows linearly at large $d$, so $v\leq (\text{const.}) \times d$ is the best thing a generic bound can do.}. Even for $J=h, d=2$, Eq.~\eqref{eq:vLRIsing} represents a 10-fold improvement over the previous bound. See Table~\ref{tab:comp} for a summary of the comparison. For the case when $h$ is large, an extension of our method will be used to derive a bound which also improves the previous results there as well, as will be shown in Sec.~\ref{sect:largeparam_TFIM}.

The LR bound for arbitrary local operators can be obtained from Eqs.~\eqref{eq:gammaBintermsofgreen} and \eqref{ineq:Zt}. For example, for $\|[\hat{\sigma}^x_{\vec{r}}(t),\hat{\sigma}^z_{\vec{0}}(0)]\|$ we have
\begin{equation}\label{eq:fourierintegralsolu}
  \|[\hat{\sigma}^x_{\vec{r}}(t),\hat{\sigma}^z_{\vec{0}}(0)]\|\leq G_{0,0}(\vec{r},t)
  =\int^\pi_{-\pi}\cosh[\omega(\vec{k})t]e^{i\vec{k}\cdot \vec{r}}\frac{d^dk}{(2\pi)^d}.
  \end{equation}

Finally, $\|[\hat{\sigma}^x_{\vec{r}}(t),\hat{\sigma}^z_{\vec{0}}(0)]\|$ also has a simple LR bound in the form of Eq.~\eqref{eq:ABfactorialbound}. Notice that in this case $d_{XY}$ is the distance between points $X=\vec{r}$ and $Y=\vec{0}$ on the commutativity graph, which is exactly $2|\vec{r}|$. Therefore, the parameter $u$ needed to satisfy Eq.~\eqref{eq:Greenfinitevelocitycond} can be chosen as $u=2v_{\mathrm{Ising}}$. Inserting Eq.~\eqref{eq:greenfactorialbound} into Eq.~\eqref{eq:fourierintegralsolu} we get the simple LR bound
\begin{equation}\label{eq:LRboundIsingsimple1}
  \|[\hat{\sigma}^x_{\vec{r}}(t),\hat{\sigma}^z_{\vec{0}}(0)]\|\leq C\left(\frac{|\vec{r}|}{v_{\mathrm{Ising}} t}\right)^{-2|\vec{r}|}, 
\end{equation}
where $C$ is a constant.

\subsubsection{Extending to arbitrary spin}\label{sec:Ising-S}
The $d$-dimensional spin-$S$ TFIM is defined by the Hamiltonian
\begin{equation}\label{eq:HIsing-S}
\hat{H}=-4J\sum_{\langle ij\rangle} \hat{S}^z_i \hat{S}^z_j-2h \sum_j \hat{S}^x_j, 
\end{equation}
where $\hat{S}^\alpha$ is the spin-$S$ operator satisfying $[\hat{S}^\alpha,\hat{S}^\beta]=i\sum_{\gamma}\epsilon^{\alpha\beta\gamma}\hat{S}^\gamma$, $\alpha,\beta,\gamma\in \{x,y,z\}$. Eq.~\eqref{eq:HIsing-S} is a direct generalization of the spin-1/2 case Eq.~\eqref{eq:tIsingH}, so in this case we can simply set $\hat{\gamma}_{\vec{r},j}=\frac{1}{S^2}\hat{S}^z_{\vec{r}}\hat{S}^z_{\vec{r}+\hat{e}_j}$ and $\hat{\gamma}_{\vec{r},0}=\frac{1}{S}\hat{S}^x_{\vec{r}}$, so that the commutativity graph for $\hat{\gamma}_{\vec{r},j},\hat{\gamma}_{\vec{r},0}$ is the same as in Fig.~\ref{fig:IsingCommu}, with $J,h$ rescaled to $4JS^2,2hS$, respectively. In this way we can borrow the result from the spin-1/2 case Eq.~\eqref{eq:vLRIsing} to get $v_{\mathrm{LR}}\leq 2X_0\sqrt{dJh}(2S)^{3/2}$. This is already qualitatively tighter than the previous best bound $v_{\mathrm{LR}}\leq 8edJ (2S)^2$.

But we can even do much better than this. Recall from the discussion at the end of Sec.~\ref{sec:upperboundHeisengerg} that the resulting LR bound would typically be tightest if we decompose $\hat{H}$ into a form where any two terms either commute or anti-commute. Now the difficulty is that in the spin-$S$ case the spin operators $\hat{S}^x,\hat{S}^z$ neither commute nor anti-commute. To overcome this, we can decompose each spin-$S$ operator as a sum of spin-1/2 Pauli operators:
\begin{equation}\label{eq:spin-Srep}
  \hat{S}^\alpha=\frac{1}{2}\sum^{2S}_{a=1}\hat{\sigma}^\alpha_a,~~~~\alpha=x,y,z,
\end{equation}
by standard addition of angular momentum. The physical Hilbert space on each site is spanned by all the eigenstates of the operator $(\hat{S}^x)^2+(\hat{S}^y)^2+(\hat{S}^z)^2$ with eigenvalue $S(S+1)$, which are simply the states that are fully symmetric under permutation of indices $a$. In the following we will derive an LR bound for operators in the enlarged Hilbert space, which automatically gives a bound on operators acting on the physical Hilbert space, since the physical operators~(e.g. $\hat{H},\hat{S}^\alpha_j$) do not couple physical states to unphysical states.

\begin{figure}
  \center{\includegraphics[width=0.45\linewidth]{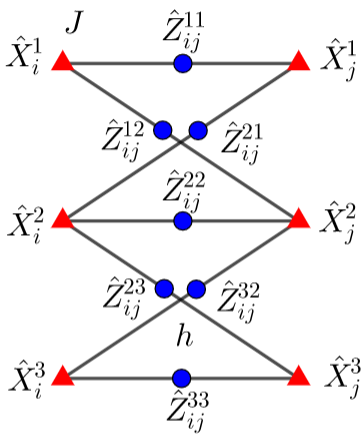}}
\caption{\label{fig:IsingCommu-S}  The commutativity graph $G$ of the spin-$S$ TFIM for the $S=3/2$ case. Only two neighboring lattice sites are drawn here. }
\end{figure}

Inserting Eq.~\eqref{eq:spin-Srep} into Eq.~\eqref{eq:HIsing-S}, the Hamiltonian in the enlarged space reads
\begin{equation}\label{eq:HIsing-S-enlarged}
\hat{H}=-J\sum_{\langle ij\rangle,1\leq a,b\leq 2S} \hat{Z}^{ab}_{ij}-h \sum_{j,1\leq a\leq 2S} \hat{X}^a_{j}, 
\end{equation}
where $\hat{Z}^{ab}_{ij}\equiv \hat{\sigma}^z_{i,a}\hat{\sigma}^z_{j,b}$, $\hat{X}^{a}_{j}\equiv \hat{\sigma}^x_{j,a}$. The commutativity graph of the enlarged Hamiltonian in Eq.~\eqref{eq:HIsing-S-enlarged} can be regarded as a decorated version of the spin-1/2 case shown in Fig.~\ref{fig:IsingCommu}, where each triangle is replaced by a set of $2S$ triangles and each circle is replace by a set of $\binom{2S}{2}$ circles, each of them linked to a distinct pair of two neighboring triangles, as shown in Fig.~\ref{fig:IsingCommu-S}. In this case Eq.~\eqref{eq:dfbart} becomes
\begin{eqnarray}\label{eq:diff_Ising-S}
  \dot{\bar{Z}}^{ab}_{ij}&=&2h(\bar{X}^a_i+\bar{X}^b_j),\nonumber\\
  \dot{\bar{X}}^a_i&=&2J\sum_{j:\langle ij\rangle,1\leq b\leq 2S}\bar{Z}^{ab}_{ij}.
\end{eqnarray}
Let $X_i\equiv\sum_{1\leq a\leq 2S}\bar{X}^a_i$ and $Z_{ij}\equiv \sum_{1 \leq a,b\leq 2S}\bar{Z}^{ab}_{ij}$, we then have
\begin{eqnarray}\label{eq:diff_Ising-S2}
  \dot{\bar{Z}}_{ij}&=&4Sh(\bar{X}_i+\bar{X}_j),\nonumber\\
  \dot{\bar{X}}_i&=&2J\sum_{j:\langle ij\rangle}\bar{Z}_{ij}.
\end{eqnarray}
This is exactly the same as the spin-1/2 case with substitution $h\to 2hS$. Therefore, borrowing the result in Eq.~\eqref{eq:vLRIsing}, we have
\begin{equation}\label{eq:vLRIsing-S}
v^S_{\mathrm{Ising}}\leq  2X_0\sqrt{2dJhS},
\end{equation}
growing as $\sqrt{S}$ at large $S$, a remarkable improvement over the previous bound which grows quadratically in $S$. Notice also that if we take the limit $S\to\infty$ at which $JS$ stays constant~(the classical limit), then our LR speed stays finite while the previous bound diverges linearly in $S$.

\subsection{The spin-$S$ Heisenberg XYZ model}\label{sec:Heisenberg-S}
The Hamiltonian for the spin-$S$ Heisenberg XYZ model is
\begin{equation}\label{eq:Heisenberg-S}
\hat{H}=\frac{2}{S}\sum_{\substack{\langle ij\rangle,\\ \alpha=x,y,z}} J_\alpha \hat{S}^\alpha_i \hat{S}^\alpha_j, 
\end{equation}
where the coefficient is normalized so that the excitation spectrum has a well-defined large-$S$ limit~\cite{haldane1983}. To get the tightest possible LR bound, we use the same trick as in Sec.~\ref{sec:Ising-S}, decomposing each spin-$S$ operator as a sum of spin-1/2 Pauli operators as done in Eq.~\eqref{eq:spin-Srep}. The enlarged Hamiltonian reads
\begin{equation}\label{eq:Heisenberg-S-enlarged}
\hat{H}=\frac{1}{2S}\sum_{\substack{\langle ij\rangle,\\1\leq a<b\leq 2S}} (J_x\hat{X}^{ab}_{ij}+J_y\hat{Y}^{ab}_{ij}+J_z\hat{Z}^{ab}_{ij}),
\end{equation}
where $\hat{X}^{ab}_{ij}\equiv \hat{\sigma}^x_{ia}\hat{\sigma}^x_{jb}$, and similarly for $\hat{Y}^{ab}_{ij},\hat{Z}^{ab}_{ij}$.
The LR bounds for the norm of the commutators $[\hat{X}^{ab}_{ij}(t),\hat{B}], [\hat{Y}^{ab}_{ij}(t),\hat{B}]$, and $[\hat{Z}^{ab}_{ij}(t),\hat{B}]$ for an arbitrary local operator $\hat{B}$ are  obtained from the solution to the differential equation~\eqref{eq:dfbart}, which in the current case becomes~(we omit the superscript $B$ for notational simplicity)
\begin{eqnarray}\label{eq:Heidiffeqn}
  S\dot{X}^{ab}_{ij}&=&\sum_{c\neq b}(J_y Y^{ac}_{ij}+J_z Z^{ac}_{ij})+\sum_{c\neq a}(J_y Y^{cb}_{ij}+J_z Z^{cb}_{ij})\\
  &&{}+\sum_{\substack{c,l\neq j:\\\langle il\rangle}}(J_y Y^{ac}_{il}+J_z Z^{ac}_{il})+\sum_{\substack{c,l\neq i:\\\langle lj\rangle}}(J_y Y^{cb}_{lj}+ J_z Z^{cb}_{lj}),\nonumber
\end{eqnarray}
where it is assumed that $i,j$ are neighboring sites, and there are two other equations obtained by permuting $X,Y,Z$. The initial condition is similar to Eq.~\eqref{eq:initialcond}. Summing over indices $a,b$, we get
\begin{equation}\label{eq:Heidiffeqnsum}
        \dot{\mathbf{X}}_{ij}=2(2-1/S)\mathbf{J}\cdot \mathbf{X}_{ij}+2 \sum_{\substack{l\neq j:\\\langle il\rangle}}\mathbf{J}\cdot \mathbf{X}_{il}+2 \sum_{\substack{l\neq i:\\\langle lj\rangle}}\mathbf{J}\cdot \mathbf{X}_{lj},
    \end{equation}
where we use the notation
\begin{equation}
  \mathbf{J}=\begin{pmatrix}
    0 & J_y & J_z \\
    J_x& 0 & J_z\\
    J_x & J_y & 0
  \end{pmatrix},
  ~~~~\mathbf{X}_{ij}=\sum_{a,b}\begin{pmatrix}
    X^{ab}_{ij}\\
    Y^{ab}_{ij}\\
    Z^{ab}_{ij}
    \end{pmatrix}.
  \end{equation}
      
  The largest eigenfrequency in a $d$-dimensional hypercubic lattice is $\omega_m(\vec{k})=4J_m[\cos\vec{k}+d-\frac{1}{2S}]$, where $J_m=\|\mathbf{J}\|$, so
  \begin{equation}\label{eq:vXYZ-S}
    v^S_{\mathrm{XYZ}}\leq 4dJ_mX_{1-\frac{1}{2Sd}}.
    \end{equation}
We now make a comparison with previous best bound and the exact solution in the $S\to\infty$ limit. For simplicity we focus on the isotropic case $J_x=J_y=J_z=J$. We have $J_m=2J$, so our LR speed remains finite $v_{\mathrm{LR}}\leq 8dJX_{1}$ in the limit $S\to\infty$, while the previous bound $v_{\mathrm{LR}}\leq 8de\|\hat{h}_{ij}\|=16deJ(S+1)$ diverges linearly. Our bound is qualitatively tight in the sense that in the large-$S$ limit, it is only a finite factor of $2X_1\approx 4.47$ bigger than the exact speed $4dJ$~(the spin-wave group velocity), as calculated from the mean-field dispersion relation in Ref.~\cite{haldane1983}. We emphasize that this is one of the few known examples of a finite LR bound for a system with an infinite local Hilbert space~\cite{cramer2008locality,nachtergaele2009lieb,schuch2011information}.

\subsection{Truncated BH model}\label{sec:truncBH}
We now consider a spin-$S$ XY model with an additional $(S+S_j^z)^2$ interaction at each site: 
\begin{equation}\label{eq:truncBH}
  \hat{H}=-\frac{J}{2S}\sum_{\langle ij\rangle}(\hat{S}^+_i\hat{S}^-_j+\mathrm{H.c.})+U\sum_j(S+\hat{S}^z_j)^2,
  \end{equation}
  where $\hat{S}^\pm=\hat{S}^x\pm i \hat{S}^y$. At each site $j$ there are $2S+1$ states $\{|m-S\rangle\}^{2S}_{m=0}$, and the action of operators $\hat{S}^\pm$ is given by
\begin{equation}
    \hat{S}^\pm|m-S\rangle 
    =\begin{cases}
      \sqrt{(m+1)(2S-m)}|m+1-S\rangle\\
      \sqrt{m(2S-m+1)}|m-1-S\rangle
       \end{cases},
\end{equation}
therefore, in the limit $S\to\infty$ the Hamiltonian in Eq.~\eqref{eq:truncBH} reproduces the BH Hamiltonian
\begin{equation}\label{eq:originalBH}
  \hat{H}=-J\sum_{\langle ij\rangle}(\hat{b}^\dagger_i\hat{b}_j+\mathrm{H.c.})+U\sum_j\hat{n}_j^2,
\end{equation}
if we identify the mapping
\begin{equation}
  \hat{S}^+\to \sqrt{2S}\hat{b}^\dagger,~~S+\hat{S}^z\to \hat{n},~~|m-S\rangle\to |m\rangle.
\end{equation}

In the following we derive an LR bound for the truncated system in Eq.~\eqref{eq:truncBH}. Using the same trick as in Sec.~\ref{sec:Heisenberg-S}, the analog of Eq.~\eqref{eq:Heidiffeqn} is~(notice that the term linear in $\hat{S}^z$ commutes with $\hat{H}$ and can therefore be dropped)
\begin{eqnarray}\label{eq:truncBHdiffeqn}
  \dot{X}^{ab}_{ij}&=&\sum_{c\neq b}\left(\frac{J}{2S} Y^{ac}_{ij}+U Z^{bc}_j\right)+\sum_{c\neq a}\left(\frac{J}{2S} Y^{cb}_{ij}+U Z^{ac}_i\right)\nonumber\\
                   &&{}+\sum_{\substack{c,l\neq j:\\\langle il\rangle}}\frac{J}{2S} Y^{ac}_{il}+\sum_{\substack{c,l\neq i:\\\langle lj\rangle}}\frac{J}{2S} Y^{cb}_{lj},\nonumber\\
  \dot{Z}^{ab}_i&=&\frac{J}{2S}\sum_{c,j:\langle ji\rangle}(X^{ac}_{ij}+Y^{ac}_{ij}+X^{bc}_{ij}+Y^{bc}_{ij}),
\end{eqnarray}
and another equation similar to the first one with $X$ and $Y$ exchanged. Summing over $a,b$ and doing a Fourier transform, we get
\begin{equation}\frac{d}{dt}\begin{pmatrix}
  X_{\vec{k}}\\
  Z_{\vec{k}}
\end{pmatrix}=\begin{pmatrix}
  2J(d+\cos\vec{k}-\frac{1}{2S}) & 2SU \\
  8J(1-\frac{1}{2S})(d+\cos\vec{k}) &0
  \end{pmatrix}\begin{pmatrix}
  X_{\vec{k}}\\
  Z_{\vec{k}}
\end{pmatrix},
\end{equation}
where $X_{\vec{k}}$ is the Fourier transform of $X_i\equiv \sum_{j:\langle ji\rangle,a,b}X^{ab}_{ij}$, and $Z_{\vec{k}}$ is the Fourier transform of $Z_i\equiv \sum_{a\neq b}Z^{ab}_{i}$. At large $S$, we have $\omega_m(i\vec{\kappa})\approx 4\sqrt{2SUJd}\cosh\frac{\kappa}{2}$, so
\begin{equation}
  v_{\mathrm{LR}}\approx 2X_0\sqrt{2SUJd}\propto\sqrt{S},
  \end{equation}
qualitatively tighter than the bound from previous methods which diverges linearly in $S$. 
It is known that the actual speed of information propagation in the $S\to\infty$ limit also diverges~\cite{eisert2009supersonic}, and mean-field approximation predicts that the BH model has $v\propto \sqrt{N}$ for an initial coherent state with average occupation number $N$~\cite{menotti2008spectral}, which suggests that the asymptotic behavior of our bound $v\propto\sqrt{S}$ is tight.

\subsection{The SU($N$) FH model}\label{sec:example_FH}
We next take the SU($N$) FH model as an example of interacting fermions, using the method in Sec.~\ref{sect:fermions}. The Hamiltonian in a general lattice is
\begin{eqnarray}\label{eq:FHH}
  \hat{H}&=&J\sum_{\langle jl\rangle,1\leq\sigma\leq N}(\hat{a}^{\dagger}_{j\sigma}\hat{a}^{\vphantom{\dagger}}_{l\sigma}+\mathrm{H.c.})\nonumber\\
  &&{}+\frac{U}{4}\sum_{j,1\leq\sigma<\sigma'\leq N}(2\hat{n}_{j\sigma}-1)(2\hat{n}_{j\sigma'}-1).
\end{eqnarray}
For $N=2$, this is the usual FH model. In the following, we limit our discussion to a bipartite lattice. We first split the fermion creation and annihilation operators to Majorana operators
\begin{equation}
  \hat{a}^\dagger_{j\sigma}=i^{(j\in E)}\frac{\hat{c}_{j\sigma}-i\hat{c}_{j\bar{\sigma}}}{2},
   \end{equation}
where $E$ is the set of even sites, $\bar{\sigma}=-\sigma$, and the Majorana operators satisfy Eq.~\eqref{eq:commuMajorana}.
The Hamiltonian in the Majorana representation is
\begin{eqnarray}\label{eq:FHHMaj}
  \hat{H}&=&\frac{J}{2}\sum_{\substack{\langle jl\rangle,j\in E,\\ 1\leq\sigma\leq N}}(i \hat{c}_{j\sigma}\hat{c}_{l\sigma}+i \hat{c}_{j\bar{\sigma}}\hat{c}_{l\bar{\sigma}})\nonumber\\
  &&{}-\frac{U}{4}\sum_{j,1\leq\sigma<\sigma'\leq N}\hat{c}_{j\sigma}\hat{c}_{j\bar{\sigma}}\hat{c}_{j\sigma'}\hat{c}_{j\bar{\sigma}'}.
\end{eqnarray}
The  norm of the operator $\hat{c}_{j\sigma;l\sigma'}(t)\equiv \{\hat{c}_{j\sigma}(t),\hat{c}_{l\sigma'}(0)\}$ is upper bounded by the solution to Eq.~\eqref{eq:dcbetat}, which in the current case become~(for notational simplicity we omit the labels $l,\sigma'$)
\begin{eqnarray}\label{eq:dcbetatFH}
  \dot{c}_{j\sigma}&=& J\sum_{l:\langle jl\rangle} c_{l\sigma}+\frac{U}{2}[c_{j\bar{\sigma}}+\sum_{\sigma'\ne \sigma}(c_{j\sigma'}+c_{j\bar{\sigma}'})],
\end{eqnarray}
with initial condition $c_{j\sigma}(0)=2\delta_{jl}\delta_{\sigma\sigma'}$. Eq.~\eqref{eq:dcbetatFH} can be easily solved by a Fourier transform. In a $d$-dimensional hypercubic lattice, we obtain the LR bound
\begin{equation}\label{eq:LRFH}
\sum_{1\leq |\sigma|\leq N}\|\{\hat{c}_{j\sigma}(t),\hat{c}_{l\sigma'}(0)\}\|\leq 2 e^{\frac{2N-1}{2}Ut}I_{\vec{x}_j-\vec{x}_l}(2Jt),
\end{equation}
where $I_\alpha(t)$ is the modified Bessel function of the first kind, and $I_{\vec{\alpha}}(t)\equiv \prod^d_{m=1}I_{\alpha_m}(t)$.
The LR speed $v_{\mathrm{FH}}$ can be calculated from Eq.~\eqref{eq:v_LR} with $\omega_m(\vec{k})=2J\cos\vec{k}+(N-1/2)U$, which gives
\begin{equation}\label{eq:vLRFH1}
  v_{\mathrm{FH}}\leq  2X_{\frac{(2N-1)U}{4dJ}}dJ.
\end{equation}
At small $U/J$, this is $3.02 dJ$, significantly improving the previous best bound $8edJN \approx 21.7 dJN$~\cite{hastings2010locality}.

At large $N$, the above LR speed scales like $N/\ln N$, which is already qualitatively tighter than the previous bound. But we can do even better than this, to prove an LR speed that grows like $\sqrt{N}$. This can either be achieved by using the commutativity graph method, or by using the fermion method with a slightly different treatment. In the following we take the latter approach for simplicity.

We go back to Eq.~\eqref{eq:FHHMaj} and write down the Heisenberg equation for the operators $\hat{c}_{j\sigma}$ and $\hat{u}_{j\sigma\sigma'}\equiv \hat{c}_{j\sigma}\hat{c}_{j\bar{\sigma}}\hat{c}_{j\sigma'}\hat{c}_{j\bar{\sigma}'}$
\begin{eqnarray}\label{eq:FHHMajHei}
  \frac{d}{dt}\hat{c}_{j\sigma}&=&J\sum_{l:\langle jl\rangle}p_{j} \hat{c}_{l\sigma}+\frac{U}{4}\sum_{\sigma'\neq\sigma}[\hat{c}_{j\sigma},\hat{u}_{j\sigma\sigma'}],\\
  \frac{d}{dt}\hat{u}_{j\sigma\sigma'}&=&\frac{J}{2}\sum_{l:\langle jl\rangle} p_{j}\nonumber\\
                                      &&{}\times[\hat{u}_{j\sigma\sigma'},\hat{c}_{j\sigma}\hat{c}_{l\sigma}+\hat{c}_{j\sigma'}\hat{c}_{l\sigma'}+\hat{c}_{j\bar{\sigma}}\hat{c}_{l\bar{\sigma}}+\hat{c}_{j\bar{\sigma}'}\hat{c}_{l\bar{\sigma}'}],\nonumber
\end{eqnarray}
where $p_{j}=(-1)^{(j\in E)}$. Following the derivations in Eqs.~(\ref{eq:gammaBt}-\ref{eq:initialcond}), one can show that the  norm of the operators $\{\hat{c}_{j\sigma}(t),\hat{c}_{l\tau}(0)\},[\hat{u}_{j\sigma\sigma'},\hat{c}_{l\tau}(0)]$ are upper bounded by the solution to the differential equations
\begin{eqnarray}\label{eq:FHHMajDE}
  \frac{d}{dt}c_{j\sigma}&=&J\sum_{l:\langle jl\rangle}c_{l\sigma}+\frac{U}{2}\sum_{\sigma'\neq\sigma}u_{j\sigma\sigma'},\\
  \frac{d}{dt}u_{j\sigma\sigma'}&=&J\sum_{l:\langle jl\rangle} [c_{j\sigma}+c_{l\sigma}+c_{j\sigma'}+c_{l\sigma'}
                                    +(\sigma,\sigma'\to\bar{\sigma},\bar{\sigma}')],\nonumber
\end{eqnarray}
with initial condition $c_{j\sigma}(0)=2\delta_{jl}\delta_{\sigma\tau},u_{j\sigma\sigma'}(0)=2\delta_{jl}(\tau\in\{\sigma,\bar{\sigma},\sigma',\bar{\sigma}'\})$. The largest eigenfrequency is
\begin{equation}\label{eq:omegaFHCommu}
\omega^{\pm}(\vec{k})=J\cos \vec{k}\pm\sqrt{J^2\cos^2\vec{k}+4UJ(N-1)(d+\cos \vec{k})}.
  \end{equation}
and the LR speed is again computed from Eq.~\eqref{eq:v_LR}. The result is $v_{\mathrm{FH}}\leq Z_{(N-1)dU/J}J$, where
\begin{equation}\label{eq:vLRFHWy}
Z_y\equiv \min_{\kappa>0}\frac{\cosh\kappa+\sqrt{\cosh^2\kappa+4y(1+\cosh\kappa)}}{\kappa}.
\end{equation}
At large $y$, $Z_y\approx X_0\sqrt{2y}$, so
\begin{equation}\label{eq:vLRFH2}
  v_{\mathrm{FH}}\approx X_0\sqrt{2NdUJ}, \text{   for large } N,
\end{equation}
a significant qualitative improvement over the previous linear growth. In the classical limit $N\to\infty$ where $\langle \hat{n}_j\rangle/N$ and $UN$ stays constant, our LR speed remains finite while previous bound diverges linearly in $N$.

\begin{figure}
    \center{\includegraphics[width=.9\linewidth]{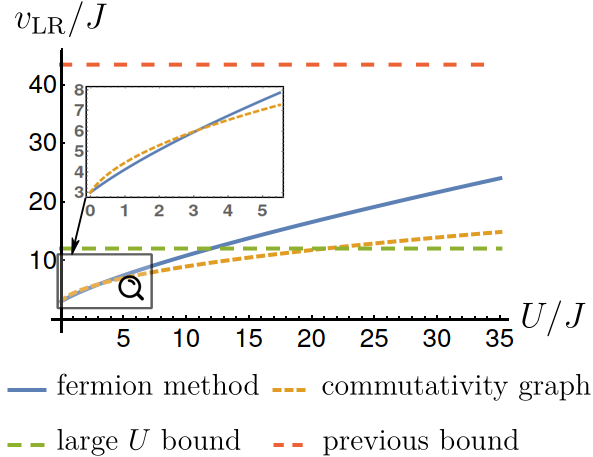}}
\caption{\label{fig:FHLRcomparison}  Comparison between different upper bounds for the LR speed of the 1D FH model, using the fermion method, the commutivity graph, and the large-$U$ method, Eqs.~\eqref{eq:dcbetat}, ~\eqref{eq:dfbart}, and~\eqref{eq:dgammaBtI}, respectively. The previous best bound is $16eJ$~\cite{hastings2010locality}. }
\end{figure}

For the 1D SU(2) case, the comparison between the LR speeds calculated from different methods is shown in Fig.~\ref{fig:FHLRcomparison}. Both methods introduced so far in this section substantially improve the previous bound for $U/J\leq 80$. Results that also improve the previous bound at large $U$ are derived in Sec.~\ref{sect:largeparam_FH}.

In the derivations above we did not use the commutativity graph method, but the method in Sec.~\ref{sec:factbound} can still be applied to derive the simple LR bound, with $d_{jl}=|\vec{x}_j-\vec{x}_l|$, and $u=\min\{2X_{\frac{(2N-1)U}{4dJ}}dJ,Z_{(N-1)dU/J}J\}$, and we have
\begin{equation}\label{eq:LRFHsimple}
\|\{\hat{c}_{j\sigma}(t),\hat{c}_{l\sigma'}(0)\}\|\leq C\left(\frac{ut}{d_{jl}}\right)^{d_{jl}},
\end{equation}
for $\forall j,l,\sigma,\sigma',t$, where $C$ is a constant.

We now compare the above bounds with the exact unequal time anti-commutator at the non-interacting point $U=0$ in a $d$-dimensional square lattice, to investigate how tight these bounds are. The Hamiltonian is diagonalized as
\begin{equation}
  \hat{H}=\int_{\mathrm{BZ}}\omega(\vec{k})\sum_{1\leq \sigma\leq N}\hat{a}^\dagger_{\vec{k}\sigma}\hat{a}^{\vphantom{\dagger}}_{\vec{k}\sigma}~d^dk,
 \end{equation}
 where $\omega(\vec{k})=2J\cos \vec{k}$. Therefore
 \begin{eqnarray}\label{eq:aisigmat}
   \hat{a}_{i\sigma}(t)&=&\frac{1}{(2\pi)^{d/2}}\int_{-\pi}^\pi \hat{a}_{\vec{k}\sigma}e^{-i\omega(\vec{k})t}e^{i\vec{k}\cdot\vec{x}}d^d k\nonumber\\
   &=&\sum_{j}J_{\vec{x}_i-\vec{x}_j}(2Jt) \hat{a}_{j\sigma}(0),
 \end{eqnarray}
where $J_n(t)$ is the Bessel function, and we define $J_{\vec{n}}(t)\equiv \prod^d_{m=1}J_{n_m}(t)$.
 Inserting Eq.~\eqref{eq:aisigmat} into the LHS of Eq.~\eqref{eq:LRFH}, we get
 \begin{equation}\label{eq:LRFHfreefermion}
\sum_{1\leq \sigma\leq N}\|\{\hat{c}_{j\sigma}(t),\hat{c}_{l\sigma'}(0)\}\|= 2 |J_{\vec{x}_j-\vec{x}_l}(2Jt)|.
\end{equation}
Comparing this with our bound in Eq.~\eqref{eq:LRFH}~[notice that the modified fermion method Eq.~\eqref{eq:FHHMajDE} also leads to the same result when $U=0$], we see that at the non-interacting point $U=0$, our LR bound just replaces the Bessel function $J_{\vec{x}_j-\vec{x}_l}(2Jt)$ of the exact solution by the modified Bessel function $I_{\vec{x}_j-\vec{x}_l}(2Jt)$. Since both $J_x(t)$ and $I_x(t)$ have the same asymptotic form $\frac{1}{x!}(\frac{t}{2})^x$ when $x/t$ is large, we conclude that our LR bound Eq.~\eqref{eq:LRFH} has the tightest large-$x$ and small-$t$ exponent at the non-interacting point, where the large-$x$ exponent $\zeta$ of a bound $f(x,t)$ is defined as  $\sup\{\zeta|\lim_{x\to \infty} e^{\zeta x\ln x}f(x,t)\text{ exists, }\forall t>0\}$, [Our bounds Eqs.~\eqref{eq:LRFH} and \eqref{eq:LRFHsimple} and the exact solution Eq.~\eqref{eq:LRFHfreefermion} all have $\zeta=1$].

\subsection{Wen's quantum rotor model}\label{sec:Wenrotor}
In this section we consider Wen's quantum rotor model~\cite{wen2003artificial,wen2003quantum,levin2006quantum}. The model is defined on a $d\geq 2$ dimensional hypercubic lattice whose sites are labeled by $\vec{r}$ and edges are labeled by $(\vec{r}\alpha)\equiv \langle \vec{r},\vec{r}+\vec{e}_\alpha\rangle$, $1\leq\alpha\leq d$. On each edge $(\vec{r}\alpha)$ there is a 2D quantum rotor, with angle variable $\hat{\theta}_{\vec{r}\alpha}$ and the dual angular momentum $\hat{S}_{\vec{r}\alpha}$, satisfying the canonical quantization condition $[\hat{\theta}_{\vec{r}\alpha},\hat{S}_{\vec{r}'\beta}]=i\delta_{\vec{r},\vec{r}'}\delta_{\alpha\beta}$. The Hamiltonian is
\begin{equation}\label{eq:Wenrotor}
\hat{H}=J\sum_{\vec{r},1\leq \alpha\leq d}\hat{S}_{\vec{r}\alpha}^2-\frac{g}{2}\sum_{\vec{r}, \alpha\neq \beta}\hat{W}_{\vec{r}\alpha\beta}, 
\end{equation}
where $\hat{W}_{\vec{r}\alpha\beta}=\exp(i\hat{\theta}_{\vec{r}\alpha}+i\hat{\theta}_{\vec{r}+\vec{e}_{\beta},\alpha}-i\hat{\theta}_{\vec{r}\beta}-i\hat{\theta}_{\vec{r}+\vec{e}_{\alpha},\beta})$. Notice that $\hat{W}_{\vec{r}\alpha\beta}=\hat{W}_{\vec{r}\beta\alpha}^\dagger$.

Based on some semiclassical treatments, Refs.~\cite{wen2003artificial,wen2003quantum,levin2006quantum} have shown that in the topologically-ordered phase $J\ll g$, this system hosts bosonic excitations propagating at speed $\sqrt{2gJ}$, behaving like artificial light waves. Ref.~\cite{Hamma2009PRL} studied a three-state truncated version of this model, and proved an LR bound with velocity $e\sqrt{2gJ}$. However, their method would fail in the untruncated version Eq.~\eqref{eq:Wenrotor}, since the term $(\hat{S}_{\vec{r}\alpha})^2$ has infinite operator norm $\|\hat{S}_{\vec{r}\alpha}^2\|=S^2$~(the number of states on a single site is $2S+1$), leading to an LR speed that diverges linear in $S$. Understanding what happens as $S\rightarrow \infty$ is important, since the argument that light emerges is valid only in this limit. In the following we use a slightly modified version of our method to overcome this difficulty, and prove that $v_{\mathrm{LR}}\leq 2X_0\sqrt{(d-1)gJ}$.

We begin by writing down the Heisenberg equations for $\hat{S}_{\vec{r}\alpha},\hat{W}_{\vec{r}\alpha\beta}$ 
\begin{eqnarray}\label{eq:WenHei}
  i\partial_t\hat{S}_{\vec{r}\alpha}&=&- \frac{g}{2}\sum_{\beta}[\hat{W}_{\vec{r}\alpha\beta}+\hat{W}_{\vec{r}-\hat{e}_\beta,\alpha\beta}-(\alpha\leftrightarrow\beta)],\nonumber\\
  i\partial_t\hat{W}_{\vec{r}\alpha\beta}&=& J\{\hat{W}_{\vec{r}\alpha\beta},\hat{S}_{\vec{r}\beta}+\hat{S}_{\vec{r}+\vec{e}_\alpha,\beta}-(\alpha\leftrightarrow\beta)\},
\end{eqnarray}
where we used $[\hat{S}_{\vec{r}\alpha},e^{\pm i\hat{\theta}_{\vec{r}'\beta}}]=\pm e^{\pm i\hat{\theta}_{\vec{r}'\beta}}\delta_{\vec{r},\vec{r}'}\delta_{\alpha\beta}$. Taking the commutator with an arbitrary local operator $\hat{B}$, we have
\begin{eqnarray}\label{eq:WenHeiB}
  i\partial_t\hat{S}^B_{\vec{r}\alpha}&=&- \frac{g}{2}\sum_{\beta}[\hat{W}^B_{\vec{r}\alpha\beta}+\hat{W}^B_{\vec{r}-\hat{e}_\beta,\alpha\beta}-(\alpha\leftrightarrow\beta)],\nonumber\\
  i\partial_t\hat{W}^B_{\vec{r}\alpha\beta}&=& J\{\hat{W}^B_{\vec{r}\alpha\beta},\hat{S}_{\vec{r}\beta}+\hat{S}_{\vec{r}+\vec{e}_\alpha,\beta}-(\alpha\leftrightarrow\beta)\}\\
                                              &&{} +J\{\hat{W}_{\vec{r}\alpha\beta},\hat{S}^B_{\vec{r}\beta}+\hat{S}^B_{\vec{r}+\vec{e}_\alpha,\beta}-(\alpha\leftrightarrow\beta)\}.\nonumber
\end{eqnarray}
Now we get rid of the first term in the second equation by doing a change of variable $\hat{W}^B_{\vec{r}\alpha\beta}(t)=\hat{U}_{\vec{r}\alpha\beta}(t)\tilde{W}^B_{\vec{r}\alpha\beta}\hat{V}_{\vec{r}\alpha\beta}(t)$, where $\hat{U}_{\vec{r}\alpha\beta}(0)=\hat{V}_{\vec{r}\alpha\beta}(0)=\hat{I}$, $i\partial_t\hat{U}_{\vec{r}\alpha\beta}(t)=J[\hat{S}_{\vec{r}\beta}(t)+\hat{S}_{\vec{r}+\vec{e}_\alpha,\beta}(t)-(\alpha\leftrightarrow\beta)] \hat{U}_{\vec{r}\alpha\beta}(t) $, and $i\partial_t\hat{V}_{\vec{r}\alpha\beta}(t)=\hat{V}_{\vec{r}\alpha\beta}(t) J [\hat{S}_{\vec{r}\beta}(t)+\hat{S}_{\vec{r}+\vec{e}_\alpha,\beta}(t)-(\alpha\leftrightarrow\beta)]$. 
Then using the same derivations in Eqs.~(\ref{eq:GammaBt2}-\ref{eq:dfbart}), we can prove that $\|\hat{S}^B_{\vec{r}\alpha}\|$ and $\|\hat{W}^B_{\vec{r}\alpha\beta}\|$ are upper bounded by the solution to the differential equation:
\begin{eqnarray}\label{eq:Wendiffeqn}
  \dot{S}_{\vec{r}\alpha}&=&g\sum_{\beta\neq \alpha}(W_{\vec{r}\alpha\beta}+W_{\vec{r}-\vec{e}_{\beta},\alpha\beta})\nonumber\\
  \dot{W}_{\vec{r}\alpha\beta}&=&2J(S_{\vec{r}\alpha}+S_{\vec{r}\beta}+S_{\vec{r}+\vec{e}_\alpha,\beta}+S_{\vec{r}+\vec{e}_\beta,\alpha}),
\end{eqnarray}
with initial condition $S_{\vec{r}\alpha}(0)= \|\hat{S}^B_{\vec{r}\alpha}(0)\|, W_{\vec{r}\alpha\beta}(0)=\|\hat{W}^B_{\vec{r}\alpha\beta}(0)\| $. The maximal eigenfrequency at $\vec{k}=i\vec{\kappa}$ is $\omega_m(i\vec{\kappa})=4\sqrt{(d-1)gJ}\cosh(\kappa/2)$, therefore, using Eq.~\eqref{eq:v_LR} we get
\begin{equation}\label{eq:v_LE_Wen1}
  v_{\mathrm{LR}}\leq 2X_0\sqrt{(d-1)gJ}.
\end{equation}
This is our second example~(after the large-$S$ Heisenberg XYZ model) of a finite LR speed in a system with an infinite local Hilbert space, and a second example~(after TFIM) of an LR speed that grows sublinearly with spatial dimension $v\propto\sqrt{d}$. It is also a good bound quantitatively--in 2D, it is only larger than the semiclassical result $v=\sqrt{2gJ}$ by a factor of $\sqrt{2}X_0\approx 2.1$~\footnote{Notice that the semiclassical result uses Cartesian distance instead of graph theoretical distance. If we use Cartesian distance as well, then the LR speed in all directions is upper bounded by  $2\sqrt{gJ}\min_{\kappa>0}\frac{\sqrt{3+\cosh\kappa}}{\kappa}\approx 2.396\sqrt{gJ}$, which is only a factor of 1.69 larger than semiclassical result.}.

\section{Improving the Lieb-Robinson velocity by eliminating large coupling constants}\label{sect:largeparam}
In several examples discussed in the previous section, our commutativity graph method dramatically improves the LR velocities when the local Hilbert space dimension or spatial dimension becomes large. Most prominently, in the large spin $S\to\infty$ limit of Heisenberg XYZ model and Wen's quantum rotor model, our method gives finite LR velocities which removes the unphysical divergence in previous bounds. Nevertheless, there is still another type of unphysical divergence present in these bounds. This happens when a Hamiltonian is a sum of terms, and one set of mutually commuting terms is multiplied by coefficients that become very large. For example,  in the large-$J$ or large-$h$ limits of TFIM in Eq.~\eqref{eq:vLRIsing}, the large-$U$ limit of SU($N$) FH model in Eqs.~\eqref{eq:vLRFH1} and \eqref{eq:vLRFH2}, and in the large-$J$ or large-$g$ limit of Wen's quantum rotor model for any finite spin $S$, the bounds for the LR velocity diverge, while in reality, the actual velocities of information propagation in these limits are expected to be finite. This kind of divergence renders the LR bound infinitely weak in these limits, a limitation that has plagued prior LR bounds.

We introduce a method, extending the results above, that removes this second type of unphysical divergence and gives a finite LR velocity in these limits. The basic idea is to derive LR bounds from coupled differential equations on a  subset of the commutativity graph that has removed the operators associated with large coefficients. This is accomplished by an appropriate unitary transform of the remaining operators. Although similar methods were employed in previous works~\cite{hastings2006,Foss2015nearly,else2020improved}, they were only used to remove single site terms, while our method can be used to remove arbitrary mutually commuting terms in the Hamiltonian. In the following we first present the general technique in Sec.~\ref{sect:largeparam_gen}, then in Sec.~\ref{sect:largeparam_TFIM} to Sec.~\ref{sect:largeparam_Wen} we apply this technique to the specific examples mentioned above, and finally present a general result in Sec.~\ref{sect:largeparam_TC}.

\subsection{Eliminating large coupling constants on the commutativity graph}\label{sect:largeparam_gen}

\begin{figure}
  \center{\includegraphics[width=\linewidth]{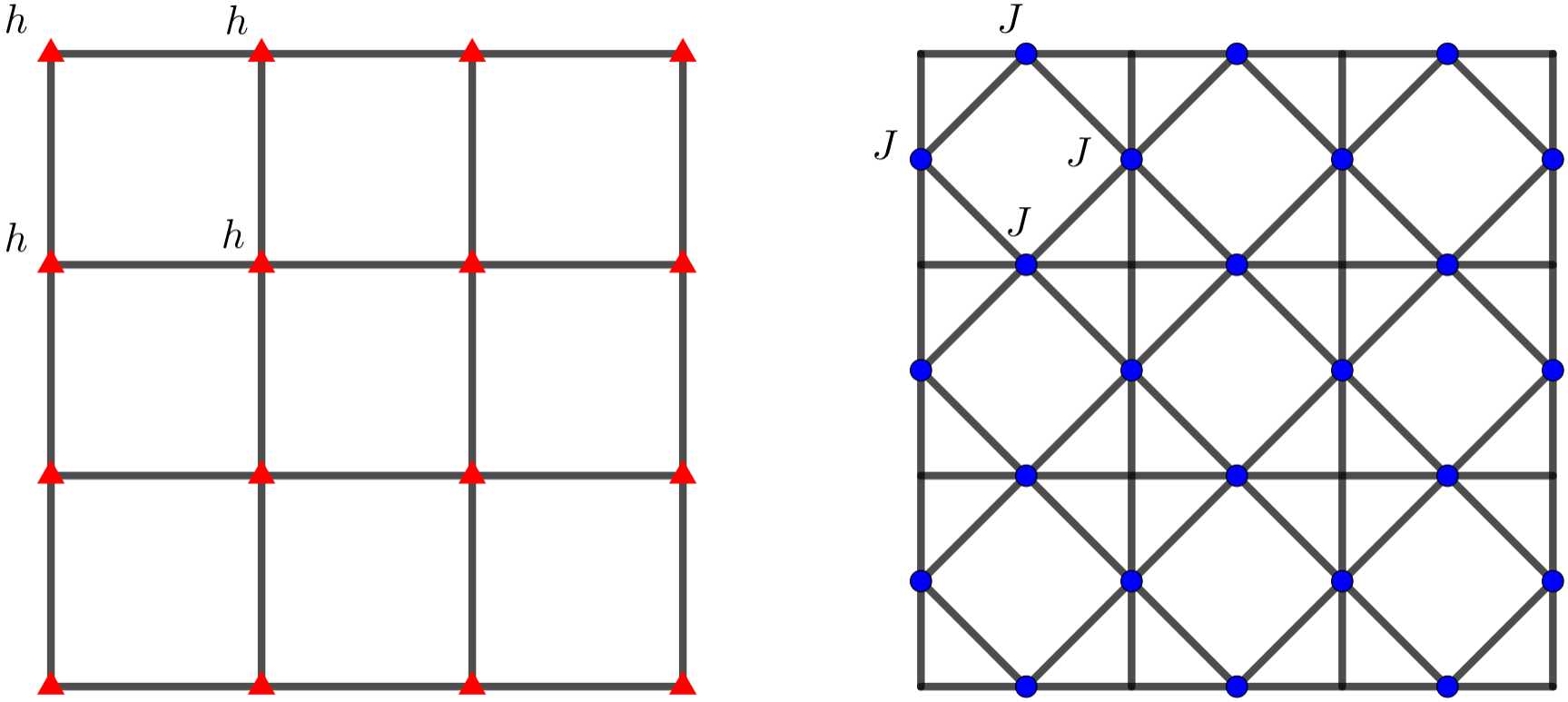}}
\caption{\label{fig:IsingCommu2} The graph $G'_F$ is constructed from the commutativity graph $G$ of the 2D TFIM in Fig.~\ref{fig:IsingCommu} by removing large parameters, by the construction in Sec.~\ref{sect:largeparam_gen}. Left: $G'_F$ for $F$ being the set of vertices representing $\hat{\sigma}^z_i\hat{\sigma}^z_j$ terms, i.e., removing parameter $J$; Right: $G'_F$ for $F$ being the set of vertices representing $\hat{\sigma}^x_i$ terms, i.e., removing parameter $h$.}
\end{figure}

Consider the commutativity graph $G=(V_G,E_G)$ of a locally-interacting spin Hamiltonian defined in Eq.~\eqref{eq:Hgeneral}, where $V_G$ is the set of vertices and $E_G$ is the set of edges.
Let $F\subset V_G$ denote a subset of disjoint vertices of $G$, such that $\{h_j |j\in F\}$ is the set of large parameters we want to get rid of. The fact that elements of $F$ are disjoint vertices in $G$ means that the corresponding operators $\{\hat{\gamma}_j|j\in F\}$ mutually commute. We now construct a reduced graph $G'_F=(V',E')$ from $G$ in the following way: (1) the vertices $V'=V_G\backslash F$ of $G'_F$ are obtained from $V_G$ by removing elements of $F$; (2) $E'$ is defined such that for $\forall i,j\in V'$, $\langle ij\rangle\in E'$ if and only if either $\langle ij\rangle\in E_G$ or there exists an $l\in F$ such that $\langle il\rangle\in E_G,\langle lj\rangle\in E_G$. In the left~(right) panel of Fig.~\ref{fig:IsingCommu2} we show the graph $G'_F$ after removing parameter $J$~($h$), respectively.

Let us denote by $\hat{I}$ the sum of all Hamiltonian terms corresponding to vertices in $F$~(\textit{i.e.} the unwanted terms):
\begin{equation}\label{eq:I_i}
  \hat{I}=\sum_{j\in F}h_j\hat{\gamma}_j.
\end{equation}
The important step here is to consider the evolution of the operator $\hat{\gamma}_i$~(for $i\in G'_F$) in a way that the unwanted  term $\hat{I}$ is ``rotated away'' by a unitary transformation:
\begin{equation}\label{eq:gammat_rotateI}
\hat{\gamma}_i(s,t)=e^{i s\hat{H}}e^{i (t-s)\hat{I}}\hat{\gamma}_ie^{-i (t-s)\hat{I}}e^{-i s\hat{H}}.
\end{equation}
Notice that $\hat{\gamma}_i(t,t)=\hat{\gamma}_i(t)$, the operator in Heisenberg picture. Taking derivative with respect to $s$, we have
\begin{eqnarray}\label{eq:dgammat_rotateI}
  \partial_s\hat{\gamma}_i(s,t)&=&e^{i s\hat{H}}i[\hat{H}-\hat{I},\hat{\gamma}_i(0,t-s)]e^{-i s\hat{H}}\nonumber\\
  &=&i\sum_{j\in G'_F,\langle ij\rangle\in G'_F}h_j [\hat{\gamma}_j(s),\hat{\gamma}_i(s,t)],
\end{eqnarray}
where the sum is over all terms in $\hat{H}-\hat{I}$ that do not commute with $\hat{\gamma}_i(0,t-s)$. Eq.~\eqref{eq:dgammat_rotateI} should be considered as the analog of Eq.~\eqref{eq:gammaHei}. Now the same derivations as in Eqs.~(\ref{eq:gammaBt}-\ref{ineq:gammaBt}) for the commutator $\hat{\gamma}^B_{i}(s,t)=[\hat{\gamma}_{i}(s,t),\hat{B}(0)]$ result in
\begin{eqnarray}\label{ineq:gammaBtI}
  \|\hat{\gamma}^B_{i}(t)\|&\leq&\|\hat{\gamma}^B_{i}(0,t)\|+ 2\sum_{\substack{j\in G'_F:\\\langle ij\rangle\in G'_F}} |h_j| \int^t_0 \|\hat{\gamma}^B_{j}(s)\|ds\nonumber\\
  &\leq& 2\|\hat{B}\|+ 2\sum_{\substack{j\in G'_F:\\\langle ij\rangle\in G'_F}} |h_j| \int^t_0 \|\hat{\gamma}^B_{j}(s)\|ds.
\end{eqnarray}
In this case the generalized Gr\"{o}nwall's inequality~\cite{Gronwall1976} indicates that $ \|\hat{\gamma}^B_{i}(t)\|$ is upper bounded by the solution
$\gamma^B_{i}(t)$ to the differential equation
\begin{equation}\label{eq:dgammaBtI}
\dot{\gamma}^B_{i}(t) = 2\sum_{j\in G'_F:\langle ij\rangle\in G'_F} |h_j|  \gamma^B_{j}(t).
\end{equation}
with initial value $\gamma^B_{i}(0)=2\|\hat{B}\|$. Eq.~\eqref{eq:dgammaBtI} is the same as Eq.~\eqref{eq:dfbart} with $G$ replaced by $G'_F$. 

In summary, when the parameters of a set of disjoint vertices of $G$ become very large, we can get a better bound for the speed of information propagation by solving the differential Eq.~\eqref{eq:dgammaBtI} on the reduced graph $G'_F$, which has the large parameters removed, and links added when two vertices of $G_F'$ are both connected to a~(removed) vertex in the original graph. Then all the methods we introduced in Sec.~\ref{sec:TI} and Sec.~\ref{sec:factbound} apply equally well. The simple form of the bound in Eq.~\eqref{eq:factorialLR} is still true provided that $d_{XY}$ and $u$ are understood as the distance and LR speed on the reduced graph $G'_F$, respectively~\footnote{Notice however, that if the methods of this section are used, then the large-$x$ and small-$t$ exponents of the resulting bounds would not be tight anymore. For example, in the TFIM, both the large-$J$ bound and the large-$h$ bound gives small-$t$ exponent $\eta(\vec{r})=|\vec{r}|$, which is looser than the bound in Sec.~\ref{sect:Ising} and exact result by a factor of 2. Since a tighter LR speed is more important in typical applications, it is generally worthwhile to tighten the LR speed at the cost of compromising the large-$x$ and small-$t$ asymptotic behaviors.}.

\subsection{Example: TFIM at large $J$ or large $h$}\label{sect:largeparam_TFIM}
As a first example, in the case of the TFIM,  the left~(right) panel of Fig.~\ref{fig:IsingCommu2} shows the graph $G_F'$ constructed from $G$ after eliminating the parameter $J$~(the parameter $h$) for the $d=2$ case, and the resulting upper bound for the LR speed is $4  X_{0}dh$~($4X_{\frac{d-1}{d}}dJ$). Therefore, combined with Eq.~\eqref{eq:vLRIsing}, we have the upper bound for the LR speed
\begin{equation}\label{eq:v_LRIsingall}
 v_{\mathrm{Ising}}\leq \min\{2X_0\sqrt{dJh},4X_{\frac{d-1}{d}}dJ,4 X_0 dh \}
\end{equation}
for the $d$-dimensional TFIM. This is a major improvement compared to the previous best bound $v_{\mathrm{Ising}}\leq 8 e dJ$~(calculated from the method in Ref.~\cite{hastings2010locality}), as we not only removed unphysical divergence of $v_{\mathrm{Ising}}$ in the large-$J$ limit, but also tightened it by a factor of at least 2.4 for all values of $d$, $J$ and $h$. [Notice however, that in 1D, a special method is available in a recent paper Ref.~\cite{Lucas2019operator}, which gives a bound for LR speed $2eJ\approx 5.44J$ for the 1D TFIM, which is about 10\% tighter than our bound $4X_0J\approx 6.04J$ in the large-$h$ limit. Away from the large-$h$ limit, when $h/J<3.24$, our bound $\min\{2X_0\sqrt{Jh},4X_0h\}$ remains the tightest. Furthermore, it is possible to combine their methods and our method of commutativity graph, which will tighten all the three velocity bounds for 1D TFIM in Eq.~\eqref{eq:v_LRIsingall} by roughly 10\%. The result will be $v_{\text{LR}}\leq \min\{e\sqrt{Jh},2eJ,2eh\}$. Notice also that if we use Theorem 4 of Ref.~\cite{Lucas2019operator} in our commutativity graph, we  may be able to  tighten our velocity bounds in higher dimensions as well. However, in $d>1$ it may be a hard combinatorical problem to compute the sum over all irreducible paths in Theorem 4 of Ref.~\cite{Lucas2019operator}, so we leave this as a future direction.]

\subsection{Example:  SU($N$) FH model at large $U$}\label{sect:largeparam_FH}
\begin{figure}
  \center{\includegraphics[width=0.7\linewidth]{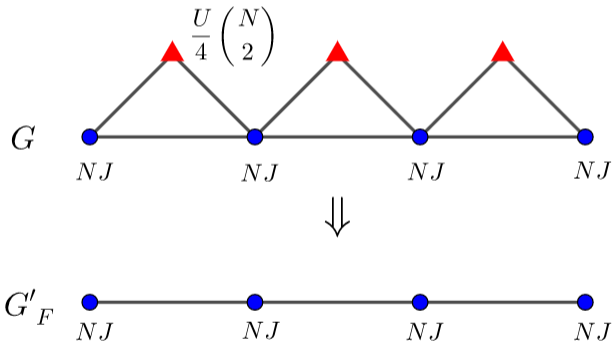}}
\caption{\label{fig:FHCommu2} Commutativity graph $G$ and the reduced graph $G'_F$ of the 1D SU($N$) FH model. Here the blue circles represent the term $\sum_{ \sigma}i \hat{c}_{j\sigma}\hat{c}_{j+1,\sigma}$ while the red triangles represent $\sum_{1\leq\sigma<\sigma'\leq N}\hat{c}_{j\sigma}\hat{c}_{j\bar{\sigma}}\hat{c}_{j\sigma'}\hat{c}_{j\bar{\sigma}'}$. $G'_F$ is obtained from $G$ by removing the red triangles. }
\end{figure}

As a second example, we consider the SU($N$) FH model. Here, the tightest large-$U$ bound is calculated from the commutativity graph $G_F'$ which is constructed from $G$ in Fig.~\ref{fig:FHCommu2} by eliminating the interaction term~(which is $\propto U$) according to the prescription of this section. The result for the LR speed is $v_{\mathrm{FH}}\leq 4X_0NJ$. Combined with the results of Sec.~\ref{sec:example_FH}, we have the upper bound
\begin{equation}
 v_{\mathrm{FH}}\leq \min\{2 X_{\frac{(2N-1)U}{4J}}J, Z_{(N-1)U/J} J,4 X_0 N J\}
\end{equation}
for the 1D SU($N$) FH model, which is also a significant improvement of the previous best bound $v_{\mathrm{FH}}\leq 8eNJ$, see Fig.~\ref{fig:FHLRcomparison} for the comparison.

\subsection{Example: Wen's  quantum rotor model at large $J$ or large $g$}\label{sect:largeparam_Wen}
As a third example of eliminating large parameters, we consider the spin-$S$ truncated version of Wen's  quantum rotor model, whose Hamiltonian has the same form as Eq.~\eqref{eq:Wenrotor}, with $\hat{S},e^{i\hat{\theta}}$~(on each site) replaced by the following $(2S+1)\times(2S+1)$ matrices:
\begin{equation}
  \hat{S}=\mathrm{diag}\{S,S-1,\ldots,-S\},~~ e^{i\hat{\theta}}=\begin{pmatrix}
    0 & I_{2S}\\
    1 & 0
    \end{pmatrix},
  \end{equation}
  where $I_{2S}$ is the $2S\times 2S $ identity matrix. Ref.~\cite{Hamma2009PRL} studied the 2D $S=1$ case, and proved an LR bound with velocity $e\sqrt{2gJ}$, in the topologically-ordered phase $J\ll g$. Here we will prove that in this limit $J\ll g$ the LR speed actually has a much tighter upper bound $v\leq \mathrm{const}.\times JS^2$. For illustrative purpose, we will focus on the 2D case for simplicity, but the generalization to arbitrary spatial dimension is straightforward. To prove this, we first notice that the commutativity graph of the Hamiltonian terms $\hat{S}_{\vec{r}\alpha}^2-S^2/2$ and $\hat{W}_{\vec{r}\alpha\beta}+\mathrm{H.c.}$ is exactly the same with that of 2D TFIM shown in Fig.~\ref{fig:IsingCommu}, where the blue circles represent the $S$ terms, with parameter $JS^2/2$, and the red triangles represent the $W$ terms, with parameter $g$.
Therefore, we can borrow the result from the TFIM with substitutions $J\to JS^2/2,~h\to g$, and we get $v_{\mathrm{LR}}\leq \min\{4X_{\frac{1}{2}}JS^2,8X_0g\}$, which is much tighter than the result of Ref.~\cite{Hamma2009PRL} $e\sqrt{2gJ}S$ in the small $J$ limit for any finite $S$. Combined with the result of Sec.~\ref{sec:Wenrotor}, we obtain
\begin{equation}\label{eq:v_LE_Wen_final}
v_{\mathrm{rotor}}\leq \min\{4X_{\frac{1}{2}}JS^2,2X_0\sqrt{(d-1)gJ},8X_0g\}.
\end{equation}

Our bound indicates that the semiclassical treatment of the rotor model would fail for any finite $S$ for sufficiently small $J$~(with fixed $g$), since the semiclassical approximation predicts emergent light traveling at speed $\sqrt{2gJ}$ which violates our bound for sufficiently small $J$. In the $S\to\infty$ limit, however, our large-$J$ bound diverges, and the method in Sec.~\ref{sec:Wenrotor} has to be used, which gives $v\leq \mathrm{const.}\sqrt{gJ}$, agreeing with the semiclassical result.

\subsection{A general result on perturbed solvable models with mutually commuting terms}\label{sect:largeparam_TC}
We end this section by making a general statement about a family of perturbed exactly solvable models, whose Hamiltonian has the form
\begin{equation}\label{eq:perturbedcommuH}
 \hat{H}=\hat{H}_0+J\sum_j\hat{V}_j,
 \end{equation}
where $\hat{H}_0$ is the solvable part and $J\sum_j\hat{V}_j$ is a sum of local perturbations. Here $j$ is a label for terms in the Hamiltonian~(it does not necessarily mean a single site: $\hat{V}_j$ can include multi-body interactions). We make the following assumptions:\\
(1) $\hat{H}_0=\sum_{i}\hat{h}_i$ is the sum of mutually commuting, locally-interacting terms, i.e. $[h_i,h_j]=0,~\forall i,j$. Primary examples include the $J$ term or the $h$ term of TFIM, the interaction term of BH model or SU($N$) FH model, the $J$ term or the $g$ term of the spin-$S$ truncated Wen's quantum rotor model, the string-net models~\cite{levin2005string} with Kitaev's toric code model~\cite{kitaev2003fault} being a prominent special case, and the exactly solvable fracton models~\cite{haah2011local,vijay2016fracton,nandkishore2019fractons}.\\
(2) There is a uniform upper bound for the norm of local perturbations $\|\hat{V}_j\|\leq c,\forall j$. This is true for any translation invariant system with a finite local Hilbert space.

Then we have
\begin{theorem}\label{thm:1}
There exist constant $c'>0$ such that $v_{\mathrm{LR}}\leq  c' |J|$, i.e. the LR speed of the model Eq.~\eqref{eq:perturbedcommuH}  grows at most linearly in the strength of the perturbation. 
\end{theorem}
Notice that the second condition is necessary, since in Wen's quantum rotor model with infinite $S$, in the limit $J\ll g$, the semiclassical treatment gives artificial light propagating at speed $v\propto \sqrt{Jg}$, which does not satisfy $v\leq c'|J|$ for sufficiently small $J$. Theorem~\ref{thm:1} can be proved by using the prescription of this section to eliminate all the terms in $\hat{H}_0$, the reduced graph $G'_{F}$ would only involve vertices representing local perturbations $\hat{V}_i$. Locality of the original Hamiltonian $\hat{H}$ implies a uniform upper bound on the degree of every vertex in $G'_F$, which, combined with condition (2), yields an LR bound with $v_{\mathrm{LR}}\leq  c' |J|$, based on similar treatment as in Eq.~\eqref{eq:appenproof1}. The previous bounds for the Hamiltonian Eq.~\eqref{eq:perturbedcommuH} give an LR speed that stays finite in the limit $J\to 0$ unless $\hat{H}_0$ only contains on-site terms~(which is not the case for models with topological order), since the previous $v_{\mathrm{LR}}$ is proportional to the operator norm of local Hamiltonians $\hat{h}_a$ in the solvable part $\hat{H}_0$. Our bound, which vanishes linearly in perturbation strength $J$, is a qualitative improvement.

\section{Consequences of tighter LR bounds on ground state correlation decay}\label{sec:appl}
Our tighter LR bounds can immediately translate into improved bounds in all areas that LR bounds are used, such as bounding the timescale for dynamical processes \cite{bravyi2006,gogolin2016equilibration,thermalization}, studying equilibrium properties like exponential decay of connected correlation functions~\cite{hastings2004decay,hastings2006,nachtergaele2006,cramer2006correlations,kliesch2014locality}, quantifying the entanglement area law~\cite{hastings2007area}, and providing error bounds on numerical algorithms~\cite{Osborne2006,Osborne2007a,Osborne2007b,woods2016dynamical,tran2019locality}. In this section we demonstrate this with one of the most important consequences of LR bounds, establishing the decay of ground state correlations in gapped, locally interacting systems~\cite{hastings2004decay,hastings2006,nachtergaele2006,cramer2006correlations}.

We will achieve this goal by directly replacing the previous LR bound by our bound Eq.~\eqref{eq:factorialLR} in the proof of Ref.~\cite{hastings2006}, and obtain a tighter upper bound on the correlation length $\xi$. Our bound is not only a quantitative improvement, but are qualitatively tighter in several limits, e.g. in the $J/h\to \infty$ limit of TFIM and the $S\to\infty$ limit of Heisenberg XXZ, where our bounds have the same asymptotic scaling as the known exact solutions.

Let us start with the following well-established inequality which relates the ground state correlation to the unequal time commutator~\cite{hastings2006,hastings2010locality}
\begin{eqnarray}\label{eq:GABG}
  |\langle G|\hat{A}_X\hat{B}_Y|G\rangle|&\leq& \frac{1}{\pi}\int^\infty_{0}\frac{e^{-\alpha t^2}}{|t|}\|[\hat{A}_X(t),\hat{B}_Y]\|dt\nonumber\\
  &&+c_0e^{-\Delta^2/4\alpha},~~~\forall\alpha>0,
\end{eqnarray}
where $c_0$ is a constant independent of $\alpha$ and $t$, and we assumed $\langle G|\hat{A}_X|G\rangle=\langle G|\hat{B}_Y|G\rangle=0$ for simplicity. We will split the integral over $t$ into two regions which we will treat differently: for $t<r/u$ we apply our LR bound in Eq.~\eqref{eq:factorialLR}, while for $t\geq r/u$ we use the trivial bound $\|[\hat{A}_X(t),\hat{B}_Y]\|\leq 2\|\hat{A}\|\|\hat{B}\|\equiv c_{1}$. We get
\begin{eqnarray}\label{eq:GABG_1}
  |\langle G|\hat{A}_X\hat{B}_Y|G\rangle|&\leq& \frac{C}{\pi}\int^{r/u}_{0}\frac{e^{-\alpha t^2}}{t} \left(\frac{ut}{r}\right)^{r} dt\\
                                         &&+\frac{c_{1}}{\pi}\int^\infty_{r/u}\frac{e^{-\alpha t^2}}{t} dt+c_0e^{-\Delta^2/4\alpha}\nonumber\\
                                           &\leq& \frac{C}{\pi}\int^{1/u}_{0}\frac{e^{-\alpha r^2 \tau^2}}{\tau} \left(u\tau\right)^{r} d\tau \nonumber\\
                                         &&+\frac{c_{1}}{\pi}\frac{e^{-\alpha r^2/u^2}}{2\alpha r^2/u^2}+c_0e^{-\Delta^2/4\alpha}.\nonumber
\end{eqnarray}
where $r=d_{XY}$, $C$ is a constant defined in Eq.~\eqref{eq:ABfactorialbound}, and in the second step we applied the inequality $e^{-\alpha t^2}\leq e^{-\alpha (r/u)^2+2\alpha (r/u) t}$ for $t\geq r/u$. If we choose $\alpha^{-1}=\lambda r$, then all three terms above decay exponentially with distance $r$, and the first term is upper bounded by $\frac{c}{\pi u} \max_{0\leq\tau\leq 1/u}e^{- r \tau^2/\lambda} u^{r}\tau^{r-1} $. Choosing $\lambda$ to maximize the smallest decay coefficient of the three terms, the lower bound for $\xi^{-1}$ is
\begin{equation}\label{eq:xibound}
  \xi^{-1}\geq \begin{cases}
\frac{\Delta}{2u}&\text{ if }\Delta\leq u\\
\frac{1}{2}W(\frac{\Delta^2e}{u^2})&\text{ if }\Delta> u,
\end{cases}
\end{equation}
where $W(x)$ is the Lambert $W$ function~(the product logarithm) defined as the solution to the equation $W e^W=x$.

\begin{figure}
    \center{\includegraphics[width=.8\linewidth]{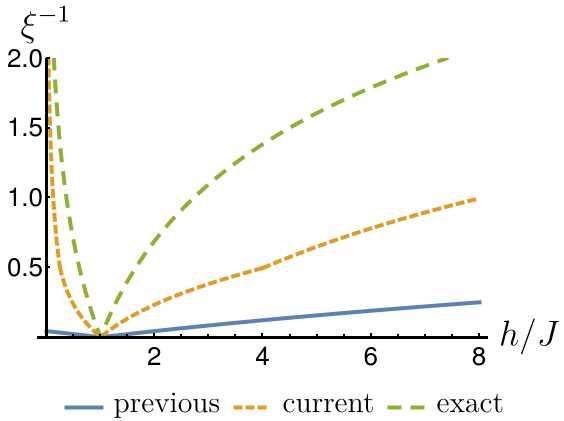}}
\caption{\label{fig:IsingGScomp} Comparison of the lower bounds for $\xi^{-1}$ in the 1D TFIM, between the result of Ref.~\cite{hastings2006} and the current paper, and the exact solution given in Ref.~\cite{pfeuty1971TFIM}. In the limit $h\gg J$, all the three curves scale as $\ln(h/J)$, while in the limit $h\ll J$, our bound and the exact solution scale as $\ln(J/h)$, while previous bound approaches a constant 0.065. }
\end{figure}

As a first specific example, we applied this bound to the 1D TFIM, and in Fig.~\ref{fig:IsingGScomp} we give a comparison between our bound with the previous best bound $\xi^{-1}\geq \max_{\mu>0}\frac{\mu}{1+16J e^\mu/\Delta}$ given in Ref.~\cite{hastings2006}, and also with the exact solution $\xi^{-1}\geq |\ln (J/h)|$ given in Ref.~\cite{pfeuty1971TFIM}~\footnote{Theorem 2 of Ref.~\cite{hastings2010locality} upper bounds the ground state correlation by $c_1\exp(\frac{r\Delta}{2v_{\text{LR}}})+c_2 g(r)$, where $c_1,c_2$ are some positive constants, and $g(r)$ first appeared in Eq.~(30) of the same paper, which, as Ref.~\cite{hastings2010locality} claims, ``decays faster than exponential in finite range interacting systems''. If this claim is correct, one would have the bound $\xi^{-1}\geq \frac{\Delta}{2v_{\text{LR}}}$, which is even tighter than our bound in Eq.~\eqref{eq:xibound} at large $\Delta/{v_{\text{LR}}}$. But this bound is actually incorrect. It is true that in locally interacting systems, the LR bound $C_B(X,t)$ decays faster than exponential at fixed time, as demonstrated in our paper. However, when the bound is written in the form Eq.~(30) of Ref.~\cite{hastings2010locality}, the function $g(r)$ only decays exponentially. This can be easily verified in the free fermion solution Eq.~\eqref{eq:LRFHfreefermion} in our paper, by noticing that the Bessel function $J_{x}(x/w)$ decays exponentially with $x$ for arbitrary $w>1$, see Ref.~\cite{watsonBessel}. Also, in the case of 1D TFIM, the bound $\xi^{-1}\geq \frac{\Delta}{2v_{\text{LR}}}$ would imply that $\xi^{-1}$ grows at least linearly with $h/J$ as $h/J$ becomes large, while the exact solution only shows a logarithmic growth, see Fig.~\ref{fig:IsingGScomp} .}. As we see, in this case our bound is always tighter than the previous bound. This is an especially strong improvement in the limit $J\gg h$ where the previous bound gives a constant, while the bound in Eq.~\eqref{eq:xibound} gives $\xi \propto \log(J/h)$, in agreement with the exact solution. 

As a second example, we apply the bound Eq.~\eqref{eq:xibound} to the large-$S$ Heisenberg XXZ model in 1D.
Our bound for the LR speed in Eq.~\eqref{eq:vXYZ-S} gives $u=v\leq 2X_1(J+\sqrt{J^2+8J_z J})$, where $J_x=J_y=J$. Therefore, our bound for $\xi$ remains finite in the limit $S\to\infty$ provided that the spectral gap $\Delta$ is finite, while the bound given by previous method~\cite{hastings2006} $\xi\leq (\text{const.})\times S/\Delta$ diverges linearly in $S$.

Finally, let us discuss the implication of Theorem~\ref{thm:1} from Sec.~\ref{sect:largeparam_TC} together with the bound in Eq.~\eqref{eq:xibound}, in the case of perturbed exactly solvable topologically ordered models with a finite local Hilbert space dimension, such as Kitaev's toric code model~\cite{kitaev2003fault},  string-net models~\cite{levin2005string}, the X-cube model~\cite{vijay2016fracton}, and Haah's cubic code model~\cite{haah2011local}. When these exactly solvable commuting Hamiltonians are perturbed with $J\sum_i\hat{V}_i$, in the limit $J\to 0$, the spectral gap $\Delta$ stays finite, while Theorem~\ref{thm:1} implies that the LR velocity on the commutativity graph has an upper bound $u\leq c'J\ll \Delta$. Therefore, Eq.~\eqref{eq:xibound} gives the bound
\begin{equation}
  \xi\leq [\frac{1}{2}W(\frac{\Delta^2e}{u^2})]^{-1}\propto [\ln\frac{\Delta}{J}]^{-1},
  \end{equation}
i.e. when the perturbation is turned off $J\to 0$, the correlation length $\xi$ vanishes at least as fast as $|\ln J|^{-1}$. The previous LR bounds for this family of perturbed exactly solvable models always give a finite LR velocity when $J\to 0$, so the previous upper bounds on $\xi$ would also be finite when $J\to 0$, qualitatively looser than our bound.

\section{Conclusions}\label{sect:summary}
We introduce a method to obtain tighter LR bounds by taking into account the details of the Hamiltonian, and show that this method improves LR bounds both qualitatively and quantitatively. The bounds established in this paper have much tighter LR speeds, especially in three broad scenarios: (1) When the number of local degrees of freedom becomes large, for example in the large-spin TFIM and Heisenberg XYZ model, the $N$-state truncated BH model, the large-$N$ SU($N$) FH model, and Wen's quantum rotor model. (2) When a set of commuting parameters becomes large, for example in the TFIM at large $J$ and the perturbed toric code model. (3) In large spatial dimension, for example in the TFIM and Wen's quantum rotor model. In these limits our LR speeds have qualitatively better asymptotic scaling than previous best bounds, and in some cases~[e.g. Wen's quantum rotor model and the classical limit of spin-$S$ Ising, Heisenberg and SU($N$) FH] our bounds are qualitatively tight in the sense that they scale in the same way as the known semiclassical solution. Our bounds also have superexponential decaying tails $e^{-d_{XY}\ln d_{XY}}$ as well as a much tighter short time scaling $O(t^{d_{XY}})$, instead of previous bounds' exponential decay $e^{-\mu d_{XY}}$ and $O(t)$ scaling. As demonstrated in several examples, the large distance and small time asymptotic behaviors of our bounds are often tightest possible.

The method introduced in this paper follows a simple prescription. To apply this method, one has to (1) Write the Hamiltonian in a suitable decomposition, ideally in a form where any two terms either commute or anticommute. (2) Draw the commutativity graph introduced in Sec.~\ref{sect:definecommu} and write down the differential equation Eq.~\eqref{eq:dfbart} whose solution gives LR bounds for the corresponding unequal time commutators; (3) Solve the differential equation. Depending on one's specific needs, we have discussed different methods to do the last step. For applications of the LR bounds in quantitative estimates, one can solve the linear differential equations numerically. In an infinite system with translation invariance, one can write the solution in a Fourier integral form, which can also be computed efficiently, and in this case we  derived a simple formula that upper bounds the LR speed. For general systems lacking translation invariance, we have derived a bound of the general form $c \left(\frac{u|t|}{d_{XY}}\right)^{d_{XY}}$, which is slightly looser than direct numerical solution to the differential equations, but is simple, much tighter than previous bounds, and retains all of the qualitative improvements offered here, namely the small-$t$ behavior, the superexponential decaying tail, and tighter LR speeds. We have illustrated our methods in a wide range of examples in Sec.~\ref{sec:examples}, where the resulting LR bounds demonstrate significant improvements over the previous best bounds. 

Our tightening of the LR bounds may have profound consequences in the study of both equilibrium and dynamical properties of quantum many body systems where the LR bounds have been applied. As a first demonstration, we showed how our bound leads to a tighter bound on the ground state correlation length in gapped systems, which qualitatively improves previous best bounds in the large-$J$ limit of the 1D TFIM and large-$S$ limit of 1D spin-$S$ Heisenberg XXZ model, and the asymptotic behavior of our bound agrees with the exact solution. Our  method to obtain much tighter LR speeds will also be useful in the quantum information context, where efforts have been made to design protocols for faster quantum state transfer~\cite{bose2007quantum,epstein2017quantum}, or schemes that create certain entangled states~\cite{clark2005efficient,monz201114,gao2013preparation}. Since the LR bounds can be used to bound the time needed for these dynamical processes~\cite{bravyi2006}, a sufficiently tight  bound not only serves as a criterion to evaluate the performance of a protocol, but also sets a theoretical upper limit on any protocols based on a given physical platform. While previous applications of the LR bounds are mostly limited to analytic proofs, the significant quantitative tightening of the bounds shown in this paper will enable numerical applications as well. One such case is to use the LR bound to upper bound the error of a local observable simulated with a finite system. Finally, as we demonstrated at the end of Sec.~\ref{sect:largeparam_Wen}, our bounds are tight enough that they may be used as a rigorous validity check for the various approximation methods used in quantum many-body physics, such as the semiclassical~(truncated Wigner) approximation~\cite{sinatra2002truncated,polkovnikov2010phase,davidson2017semiclassical}, mean-field approximations~\cite{altland2010condensed,kamenev2011field}, and random phase approximation~\cite{negele2018quantum}. An approximation is immediately proved invalid if it predicts a propagation speed $v$ that exceeds our bound, or if it gives ground state correlation length $\xi$ and spectral gap $\Delta$ that does not satisfy the inequality in Eq.~\eqref{eq:xibound}.

The methods introduced in this paper may also shed light on some of the important unsolved problems in the area of LR bounds. We mention two of them below.

One important open problem is how to extend LR bounds to systems with an infinite local Hilbert space dimension, such as in systems with interacting bosons. Despite extensive effort, up to now there are only some preliminary success in obtaining a finite LR bound in harmonic systems~\cite{cramer2008locality,nachtergaele2009lieb}, or bounds for a restricted set of operators and special initial states~\cite{schuch2011information} in BH-type models. Our method is very promising in this direction, as already demonstrated in this paper, it leads to a finite LR bound for large-$S$ Heisenberg XYZ model and Wen's quantum rotor model, and has shown qualitatively better scaling laws of $v_{\mathrm{LR}}$ with number of local states in truncated BH model and SU($N$) FH model, all of which have not been achieved before. In future, we hope that by extending the methods of this paper, one can find a bigger class of nontrivial interacting models with infinite local Hilbert space whose LR speeds remain finite. 

Another open problem is to tighten the LR bounds in systems with power-law decaying interactions~\cite{Foss2015nearly}. The generalization of our method to this case is straightforward. It may not lead to a qualitative improvement in the most generic case, but when applied to specific models, our method can take advantage of the specific properties of the Hamiltonian to give improved bounds. A promising case is when the long-range interactions are mutually commuting. This includes Coulomb interactions, van der Waals interactions between Rydberg atoms, the $\sigma^z_i\sigma^z_j$ interaction in power-law TFIM, all of which are of great theoretical and experimental interests. In Sec.~\ref{sec:examples} we have seen in many examples that our commutativity graph method has a qualitative advantage over previous methods especially when there are lots of commuting terms that have a spatial overlap. It is therefore natural to expect that if one combines our commutativity graph method with the current best methods~\cite{tran2019locality,Lucas2019longrange,kuwahara2020strictly} in treating the long-range part, one can get qualitatively better, model-specific LR bounds in a family of models that are of great physical significance.


\acknowledgements
We thank Stefano Chessa, Yichen Huang, Andrew Lucas, Michael Foss-Feig, Putian Li, Bhuvanesh Sundar, and Ian White for discussions. This work was supported by the Welch Foundation (C-1872) and the National Science Foundation (PHY-1848304).

\appendix
\section{A more general method}\label{appen:unifiedapproach}
In the examples of Sec.~\ref{sec:examples}, there are two places where our method slightly deviates from the general methods introduced in Sec.~\ref{sect:generalmethod}: in the SU($N$) FH model Eqs.~(\ref{eq:FHHMajHei}-\ref{eq:FHHMajDE}) we used a method which is essentially a mixture of the commutativity graph method and the fermion method, and in Wen's quantum rotor model Eq.~\eqref{eq:WenHei} we started with the Heisenberg equation for $\hat{S}^z_{ij}$ and $\hat{W}_i$ rather than the Hamiltonian terms $(\hat{S}^z_{ij})^2$ and $\hat{W}_i$ as we did in Sec.~\ref{sect:definecommu}. One may therefore be curious whether there is a more general prescription behind these seemingly different approaches. Indeed, as we will see in this section, the commutativity graph method, the fermion method, the modified fermion method in the SU($N$) FH model, and the method for Wen's quantum rotor model, are just special cases of a more general method to be introduced below. This general method  starts with the Heisenberg equation for an arbitrary set of local operators, thereby allowing a high level of flexibility. The downside of this flexibility is that in some cases one may need considerable physical and mathematical intuition to make the best choice. The general prescription is:

Step 1. Choose a local operator basis $\{\hat{\Omega}_i\}_{i\in M}$ for the Hamiltonian $\hat{H}$, here $M$ is a collection of indices. By basis, we mean that the Hamiltonian can be written as a sum of products of elements of this basis in at least one way~(there can be more than one way, in this case we call the basis overcomplete)
\begin{equation}\label{eq:unifiedH}
  \hat{H}=\hat{H}[\{\hat{\Omega}_i\}]=\sum_{I}h_I\prod_{i\in I}\hat{\Omega}_i,
\end{equation}
where $I$ denotes an ordered subset of $M$. For example, in the commutativity graph method, the basis is simply the collection of all individual terms in the Hamiltonian Eq.~\eqref{eq:Hgeneral}; in the fermion method, the basis is the set of all Majorana operators; in the modified fermion method of FH model, the basis is the set of all Majorana operators and all the $\hat{u}_{j\sigma\sigma'}$ terms in Eq.~\eqref{eq:FHHMajHei}, however, since $\hat{u}_{j\sigma\sigma'}$ is a quartic product of Majorana operators, this basis is overcomplete; in Wen's quantum rotor model, the basis is the collection of all $\hat{S}^z_{ij}$ and $\hat{W}_i$. One can easily verify that in all these cases, the Hamiltonian can be written as a polynomial of the basis elements.

Step 2. Write down the Heisenberg equation of motion for the basis operators, as is done in Eqs.~(\ref{eq:gammaHei}, \ref{eq:dcFHt}, \ref{eq:FHHMajHei}, \ref{eq:WenHei}). For the generic Hamiltonian in Eq.~\eqref{eq:unifiedH}, we have
\begin{eqnarray}\label{eq:unifiedHei}
  i\partial_t\hat{\Omega}_i&=&\sum_{I}h_I[\hat{\Omega}_i,\prod_{j\in I}\hat{\Omega}_j],\nonumber\\
                           &=& \sum_{J}h'_J\prod_{l\in J}\hat{\Omega}_l,\\
                           &=&   P_i[\{\hat{\Omega}_j\}]\hat{\Omega}_i+\hat{\Omega}_iQ_i[\{\hat{\Omega}_j\}]+R_i[\{\hat{\Omega}_j\}],\nonumber
  \end{eqnarray}
 where in the second line we apply the basic commutation relations of the basis operators to simplify the equation, and in the third line we isolate all the terms that are left-proportional or right-proportional to $\hat{\Omega}_i$, here $P_i,Q_i,R_i$ are some polynomial functions of the basis operators, and $P_i[\{\hat{\Omega}_j\}], Q_i[\{\hat{\Omega}_j\}]$ are assumed to be Hermitian~(we can always absorb the non-Hermitian part into $R_i[\{\hat{\Omega}_j\}]$). 

Step 3. On both sides of the Heisenberg equation, take the commutator with another local operator $\hat{B}$ (note: if both the LHS and $\hat{B}$ are fermionic operators, anti-commutator should be taken). Use Leibniz's rule to expand the product terms, e.g. $[\hat{a}\hat{b}\hat{c},\hat{B}]=[\hat{a},\hat{B}]\hat{b}\hat{c}+\hat{a}[\hat{b},\hat{B}]\hat{c}+\hat{a}\hat{b}[\hat{c},\hat{B}]$. This is done in Eqs.~(\ref{eq:gammaBt}, \ref{eq:dcbetatgeneral}, \ref{eq:WenHeiB}) [and below Eq.~\eqref{eq:FHHMajHei}, implicitly] of the main text. We write the equation in the form
 \begin{equation}\label{eq:unifiedHeiB}
  i\partial_t\hat{\Omega}^B_i=  P_i[\{\hat{\Omega}_j\}]\hat{\Omega}^B_i+\hat{\Omega}^B_iQ_i[\{\hat{\Omega}_j\}]+S_i[\{\hat{\Omega}_j\},\{\hat{\Omega}_j^B\}],
  \end{equation}
where $S_i[\{\hat{\Omega}_j\},\{\hat{\Omega}_j^B\}]$ is some polynomial function that is linear in $\hat{\Omega}_j^B$.

Step 4. Use a change of variable  $\hat{\Omega}^B_i(t)\to \hat{U}_i(t) \hat{X}^B_i(t) \hat{V}_i(t)$ to cancel the first two terms on the RHS of Eq.~\eqref{eq:unifiedHeiB}, where $i\partial_t\hat{U}_i(t)=P_i[\{\hat{\Omega}_j\}] \hat{U}_i(t), i\partial_t\hat{V}_i(t)=\hat{V}_i(t) Q_i[\{\hat{\Omega}_j\}]$~\footnote{One can actually get rid of more general terms of the form $\sum_a \hat{P}_a  \hat{\Omega}^B_i \hat{Q}_a$ for arbitrary Hermitian operators $\hat{P}_a,\hat{Q}_a$~(which are implicitly functions of the basis operators $\{\hat{\Omega}_j\}$) on the RHS of Eq.~\eqref{eq:unifiedHeiB}, i.e. on the RHS of Heisenberg equation for $i\partial_t\hat{\Omega}^B_i$ one can get rid of any product term that is proportional to $\hat{\Omega}^B_i$ with coefficients being Hermitian operators. This can be done by going to the Liouville picture, where Eq.~\eqref{eq:unifiedHeiB} can be rewritten as $i\partial_t |\Omega^B_i)= \sum_a\hat{\Pi}_a |\Omega^B_i) +\sum_{j\neq i} \hat{S}_j |\Omega^B_j)$, where the operator $\hat{\Omega}_j^B$ is mapped to a state $|\Omega^B_i)$, $\hat{\Pi}_a$ denotes the superoperator representing the map $\hat{\Omega}^B_i \to \hat{P}_a \hat{\Omega}^B_i\hat{Q}_a$, and $\hat{S}_j$ denotes the superoperator representing the map $\hat{\Omega}^B_j\to S_i[\{\hat{\Omega}_j\},\{\hat{\Omega}^B_j\}]$~(remember that $S_i[\{\hat{\Omega}_j\},\{\hat{\Omega}^B_j\}]$ is linear in $\hat{\Omega}^B_j$). If $\hat{P}_a,\hat{Q}_a$ are all Hermitian, then $\hat{\Pi}_a$ is Hermitian, and the first term can also be rotated away by a unitary rotation in the Liouville picture, which also preserves the norm of the state $|\Omega^B_i)$. The steps 5 and 6 will follow without problem.}. Eq.~\eqref{eq:unifiedHeiB} becomes
 \begin{equation}\label{eq:unifiedHeiBUV}
  i\partial_t\hat{X}^B_i=  \hat{U}^\dagger_i S_i[\{\hat{\Omega}_j\},\{\hat{\Omega}_j^B\}]\hat{V}^\dagger_i.
  \end{equation}
Notice also $\|\hat{\Omega}^B_i(t)\|=\|\hat{X}^B_i(t)\|$ due to $\hat{U}_i,\hat{V}_i$ being unitary. This is done in Eq.~\eqref{eq:GammaBt2} [and below Eq.~\eqref{eq:FHHMajHei}, implicitly]  and between Eqs.~(\ref{eq:WenHeiB}, \ref{eq:Wendiffeqn}).

Step 5. Take operator norm on both sides of the resulting operator evolution equation~\eqref{eq:unifiedHeiBUV}, and apply the triangle inequalities on the RHS to get $\|S_i[\{\hat{\Omega}_j\},\{\hat{\Omega}_j^B\}]\|\leq \sum_j S_{ij}\|\hat{\Omega}_j^B\|$ for suitable non-negative coefficients $S_{ij}$.  Then using the fundamental theorem of calculus and Gr\"{o}nwall's inequality, one can prove that $\|\hat{\Omega}^B_i(t)\|$ is upper bounded by the solution to the system of linear differential equations
\begin{equation}\label{eq:unifieddiffeqn}
  \partial_t\bar{\Omega}^B_i=   \sum_j S_{ij}\bar{\Omega}_j^B
  \end{equation}
with initial condition $\bar{\Omega}^B_i(0)=\|\hat{\Omega}^B_i(0)\|$. This is done in Eqs.~(\ref{ineq:gammaBt}-\ref{eq:dfbart}, \ref{ineq:dcbetat}-\ref{eq:dcbetat}, \ref{eq:FHHMajDE}, \ref{eq:Wendiffeqn}).

Step 6. Solve the resulting linear differential equations~\eqref{eq:unifieddiffeqn}, and calculate $v_{\mathrm{LR}}$ from its maximal eigenfrequency using Eq.~\eqref{eq:v_LR}.

The main flexibility allowed by this general method is the choice of basis operators $\{\hat{\Omega}_i\}_{i\in M}$. A different choice of basis would typically lead to a different LR speed. We would hope to find a basis that makes the resulting LR speed as small as possible to make the LR bound tighter. We have already demonstrated in the main text that the basis $\{\hat{\gamma}_i\}$ used in the minimal Clifford decomposition Eq.~\eqref{eq:Hgeneral} satisfying Eq.~\eqref{eq:opalg} is a good choice of basis for a large class of spin and interacting fermion models, and the Majorana operator basis $\{\hat{c}_i\}$ satisfying Eq.~\eqref{eq:commuMajorana} is good for weakly interacting fermions, in the sense that the resulting LR speeds vastly improves the previous results, and the LR bounds typically have tight large distance and small time exponents. We did encounter a few exceptions where other choices are favorable. For example in Wen's quantum rotor model we used the basis $\{\hat{S}^z_{ij},\hat{W}_i,\hat{W}^\dagger_i\}$, and in the large-$N$ SU($N$) FH model we used $\{\hat{c}_{i\sigma},\hat{u}_{j\sigma\sigma'}\}$. Also, in the spin-$S$ Heisenberg model, an alternative, simpler choice of basis $\{S^x_i, S^y_i,S^z_i\}$ leads to $v\leq 4dJ_m X_1$, which is the same as Eq.~\eqref{eq:vXYZ-S} in the $S\to\infty$ limit. It may be a good future direction to further explore the role of basis operators and hopefully obtain a general prescription on how to find the best choice of basis in a specific model.

\bibliography{LR}

\end{document}